\documentclass[journal]{IEEEtran}
\usepackage{tikz}
\usepackage[cmex10,intlimits]{amsmath}
\usepackage{amsfonts}
\usepackage{amssymb}

\usepackage[european resistors]{circuitikz}
\usetikzlibrary{calc}

\usepackage{capekCommands}
\usepackage{capekColors}
\graphicspath{figures/}
\usepackage[normalem]{ulem}

\providecommand{\RmatMat}{\ensuremath{\M{R}_\rho}}
\providecommand{\RmatVac}{\ensuremath{\M{R}_\srcRegion}}
\providecommand{\XmatVac}{\ensuremath{\M{X}_\srcRegion}}

\providecommand{\DIMmat}{\ensuremath{\M{D}}}
\providecommand{\PORTmat}{\ensuremath{\M{C}}}

\providecommand{\Vport}{\ensuremath{\M{v}}}
\providecommand{\Iport}{\ensuremath{\M{i}}}
\providecommand{\Fport}{\ensuremath{\M{f}}}

\providecommand{\Yport}{\ensuremath{\M{y}}}
\providecommand{\YportL}{\ensuremath{\Yport_\T{L}}}
\providecommand{\ZcharMat}{\ensuremath{\V{\Lambda}}}
\providecommand{\KportMat}{\ensuremath{\M{k}_\T{i}}}
\providecommand{\LportMat}{\ensuremath{\M{k}_\T{r}}}
\providecommand{\Gport}{\ensuremath{\M{g}_\srcRegion}}
\providecommand{\Lport}{\ensuremath{\M{g}_\rho}}

\providecommand{\TARC}{\ensuremath{\Gamma^\T{t}}}
\providecommand{\Grealized}{\ensuremath{G}^\T{t}}

\providecommand{\Ptot}{\ensuremath{P_\T{in}}}

\providecommand{\apar}{\ensuremath{\M{a}}}
\providecommand{\bpar}{\ensuremath{\M{b}}}

\begin{document}

\title{Finding Optimal Total Active Reflection Coefficient and Realized Gain for Multi-port Lossy Antennas}
\author{Miloslav~Capek, \IEEEmembership{Senior Member, IEEE}, Lukas~Jelinek, and Michal~Masek
\thanks{Manuscript received \today; revised \today. This work was supported by the Ministry of Education, Youth and Sports under project~\mbox{LTAIN19047}, by the Czech Science Foundation under project~\mbox{No.~19-06049S}, and by the Grant Agency of the Czech Technical University in Prague under grant \mbox{SGS19/168/OHK3/3T/13}.}
\thanks{M. Capek, L. Jelinek, and M. Masek are with the Czech Technical University in Prague, Prague, Czech Republic (e-mails: \{miloslav.capek; lukas.jelinek; michal.masek\}@fel.cvut.cz).}
\thanks{Color versions of one or more of the figures in this paper are
available online at http://ieeexplore.ieee.org.}
\thanks{Digital Object Identifier XXX}
}

\maketitle

\begin{abstract}
A numerically effective description of the total active reflection coefficient and realized gain are studied for multi-port antennas. Material losses are fully considered. The description is based on operators represented in an entire-domain port-mode basis, \ie{}, on matrices with favorably small dimensions. Optimal performance is investigated and conditions on optimal excitation and matching are derived. The solution to the combinatorial problem of optimal ports' placement and optimal feeding synthesis is also accomplished. Four examples of various complexity are numerically studied, demonstrating the advantages of the proposed method. The final formulas can easily be implemented in existing electromagnetic simulators using integral equation solver.
\end{abstract}


\section{Introduction}

\IEEEPARstart{W}{ith} increasing interest in multi-port and MIMO systems, see, \eg{},~\cite{Manteuffel_Martens-CompactMultimodeMultielementAntennaForIndoorUWB, RazaYangHussain_MeasurementOfRadEff_2012, LiLauYingHe_CharacteristicModeBasedTradeoffAnalysis, Hanulla_etal_FrequencyReconfiguragleMultibandHandset, Luomaniemi_SwitchReconfigurableMetalRimMIMOhandset, Saarinen_etal_CombinatoryFeedingMethodForMobileApplication}, a new set of figure of merits were needed to judge their performance. Among the few available choices, the total active reflection coefficient (TARC)~\cite{2003_ManteghiRahmatSamii_TARC, 2005_ManteghiRahmatSamii_TARC, Jamaly_Derneryd_EfficiencyCharacterisationOfNport} became widely used, due mainly to its straightforward definition and experimental accessibility. Considering a given excitation, it reads \Quot{the square root of the available power generated by all excitations minus radiated power, divided by available power} \cite{2005_ManteghiRahmatSamii_TARC}. 

The TARC was successfully utilized for practical antenna design, covering wide range of MIMO applications \cite{Browne_etal_ExperimentsWithCompactAntennaArraysForMIMORadioCom, Chae_etal_AnalysisOfMutualCouplingCorrelationsAndTARCinWiBroMIMOArrayAntenna, Costa_etal_EvaluationOfANewWidebandSlotArrayForMIMOPerformance, Hannula_etal_ConceptForFrequencyReconfigurableAntennaBasedOnDistributedTransceivers}. Improvement of large-scale end-fire antenna arrays matching with simultaneous pattern synthesis was achieved via convex optimization in ~\cite{Helander_etal_SynthesisOfLargeEndfireAntennaArraysUsingConvexOpt}. The optimal performance on TARC in a loss-less scenario using scattering matrix was studied in~\cite{Hanulla_etal_FrequencyReconfiguragleMultibandHandset}.

Numerical evaluations of TARC typically neglect ohmic losses (relating TARC to matching efficiency only), assuming that vanishing TARC implies the acceptance of all incident power by the antenna and radiation of it into far field~\cite{Hannula_etal_TunableEightElementMIMOAntennaBasedOnTheAntennaClusterConcept}. This assumption is, nevertheless, not true when ohmic losses are present and it may be a source of significant discrepancies between simulation and measurement~\cite{Wang_etal_AClosedFormFormulaOfRadiationAndTotalEfficiencyForLossyMultiport}. In order to remedy this issue and provide simple, yet precise TARC evaluation, a new formulation has been devised in this paper (relating TARC to the total efficiency). Unlike other procedures relying on equivalent circuits~\cite{Jamaly_Derneryd_EfficiencyCharacterisationOfNport, Wang_etal_AClosedFormFormulaOfRadiationAndTotalEfficiencyForLossyMultiport}, the presented approach is of full-wave nature, \ie{}, there are many discrete ports which can be freely distributed across arbitrarily shaped scatterer, optionally made of lossy and inhomogennous material. Another notable feature is that characteristic impedance can be separately prescribed for each port.

The presented derivations make use of port modes~\cite{1978_Harrington_TAP, Harrington_FieldComputationByMoM}, which compress the large algebraic system (typically thousands times thousands) describing the antenna into port-related matrix operators of rank given by the number of ports (typically not more than tens by tens). Due to this property, the evaluation of antenna metrics is numerically inexpensive and allows for optimal design. Particularly, assuming a given shape, materials, operational frequency, number of ports and matching topology, this manuscript introduces methodology, how to determine where the ports have to be placed, what is the best combination of excitation voltages and what are the optimal matching impedances to reach the best attainable values of TARC and realized gain. The understanding provided by the method also updates the knowledge about optimal excitation of antenna arrays~\cite{1978_Harrington_TAP, HarringtonMautz_ControlOfRadarScatteringByReactiveLoading, HarringtonMautz_PatternSynthesisForLoadedNportScatterers, Bucci1994, Hazdra_etal_OnEndFireSuperDirectivityOfArrays}.

This paper is structured as follows. TARC is briefly reviewed and expressed in operator form in Section~\ref{sec:tarc}, and reformulated in port modes in Section~\ref{sec:PMs}. The fundamental bounds on TARC performance are derived in Section~\ref{sec:TARCbound}. To demonstrate the usefulness of the novel formula and its capability of determining optimal performance of a multi-port antenna, the optimal placement of feeding ports is shown in Section~\ref{sec:synthesis} and optimal characteristic impedances are found in Section~\ref{sec:terminImpend}. The realized gain and its fundamental bound are treated in Section~\ref{sec:realGain}. The paper is concluded in Section~\ref{sec:concl}. All relevant sections are accompanied with numerical examples demonstrating flexibility of the method.

\section{TARC -- Full-wave Algebraic Formulation}
\label{sec:tarc}

\begin{figure}
\centering
\includegraphics[width=\columnwidth]{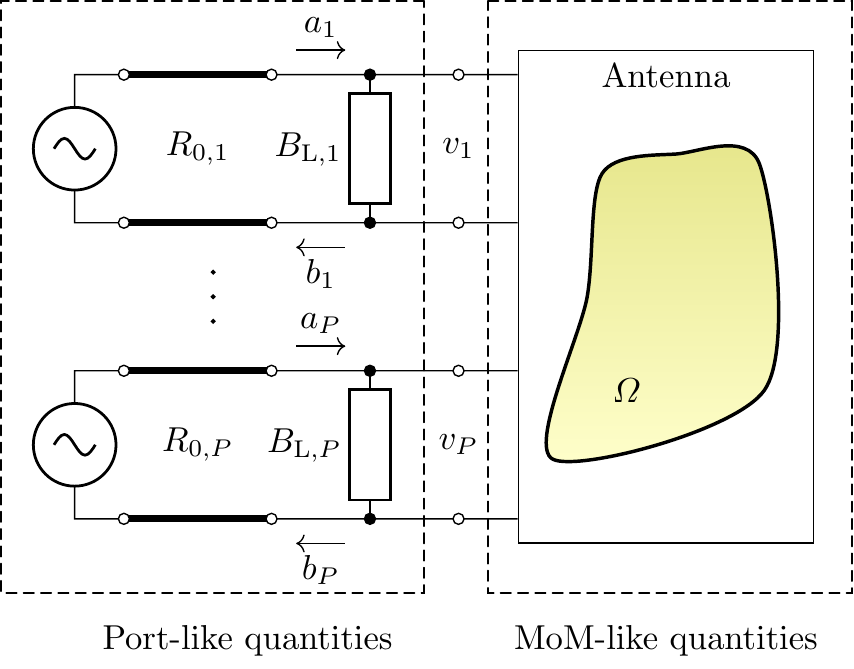}
\caption{Multi-port antenna system, consisting of scatterer~$\srcRegion$ and $P$~ports connected to transmission lines of characteristic impedance~$R_{0,p}$. All ports might optionally be tuned by a lumped susceptance~$B_{\T{L},p}$. The variable connecting circuitry on the left and the full-wave model on the right are port voltages accumulated in vector~$\Vport$.}
\label{fig:fig1}
\end{figure}

Considering a multi-port antenna as depicted in Fig.~\ref{fig:fig1}, the total active reflection coefficient (TARC) is defined as~\cite{2005_ManteghiRahmatSamii_TARC}
\begin{equation}
\TARC = \sqrt{1 - \dfrac{\Prad}{\Ptot}},
\label{eq:TARCdef}
\end{equation}
where~$\Prad$ is the power radiated by the antenna and $\Ptot$ is the incident (available) power. The incident power is most easily evaluated as
\begin{equation}
\Ptot = \dfrac{1}{2} \apar^\herm \apar,
\label{eq:Ptotdef}
\end{equation}
where~$\apar \in \mathbb{C}^{P \times 1}$ is a vector of incident power waves~\cite{Pozar_MicrowaveEngineering} with $P$ being the number of ports and superscript~$\herm$ denoting the Hermitian conjugate. In the case of a loss-less antenna, the radiated power can be evaluated as
\begin{equation}
\Prad = \dfrac{1}{2} \left(\apar^\herm \apar - \bpar^\herm \bpar \right),
\label{eq:Praddef1}
\end{equation}
where~$\bpar \in \mathbb{C}^{P \times 1}$ is the vector of reflected power waves. For a lossy antenna, this is no longer valid. In a general case, the radiated power must be evaluated as~\cite{Harrington_FieldComputationByMoM}
\begin{equation}
\Prad = \dfrac{1}{2} \Ivec^\herm \RmatVac \Ivec,
\label{eq:Praddef2}
\end{equation}
where $\RmatVac\in \mathbb{R}^{N\times N}$ is the radiation part of the impedance matrix~\cite{Harrington_FieldComputationByMoM}
\begin{equation}
\Zmat = \RmatVac + \RmatMat + \J \XmatVac,
\label{eq:Zmatdef}
\end{equation}
and $\Ivec\in\mathbb{C}^{N\times 1}$ is a vector of expansion coefficients within a method of moments (MoM) solution~\cite{Harrington_FieldComputationByMoM} to surface current density
\begin{equation}
\V{J} \left(\V{r}\right) \approx \sum_{n=1}^N I_n \basisFcn_n \left(\V{r}\right)
\label{eq:Jvecdef}
\end{equation}
with~$\left\{\basisFcn_n\right\}$ being a properly chosen set of basis functions~\cite{PetersonRayMittra_ComputationalMethodsForElectromagnetics}. The power dissipated in ohmic losses\footnote{If volumetric method of moments is used, matrix~$\M{X}_\rho$ would have to be added to \eqref{eq:Zmatdef} to account for the presence of dielectric bodies. The rest of the formulation remains untouched.} is correspondingly evaluated as
\begin{equation}
\Plost = \dfrac{1}{2} \Ivec^\herm \RmatMat \Ivec,
\label{eq:Plostdef}
\end{equation}
where matrix~$\RmatMat\in \mathbb{R}^{N\times N}$ is calculated for a surface resistivity model as described in~\cite[App.~C]{JelinekCapek_OptimalCurrentsOnArbitrarilyShapedSurfaces}. The procedure of how to evaluate current~$\Ivec$ with connected lumped susceptances~$\left\{B_{\T{L,p}}\right\}$ is described in the following section.

The substitution of~\eqref{eq:Ptotdef} and~\eqref{eq:Praddef2} into~\eqref{eq:TARCdef} is, in principle, enough to evaluate TARC, nevertheless, such a prescription is unpleasant when mixing port-based ($\apar$, $\bpar$) and MoM-based ($\Ivec$) quantities. This form, as an example, does not allow for evaluating optimal performance~\cite{GustafssonTayliEhrenborgEtAl_AntennaCurrentOptimizationUsingMatlabAndCVX, JelinekCapek_OptimalCurrentsOnArbitrarilyShapedSurfaces, CapekGustafssonSchab_MinimizationOfAntennaQualityFactor}, which will be derived later. In order to overcome this difficulty, the TARC formula will now be recast into a form which solely contains port quantities.

\section{Expression of TARC in Port Quantities}
\label{sec:PMs}
In order to rewrite TARC in terms of port-related quantities, \ie{}, matrices and vectors of size~$P$, the excitation vector~$\Vvec$ and current vector~$\Ivec$ must be related to port voltages~$\Vport$ and port currents~$\Iport$.

In the first step, excitation vector~$\Vvec$ from the MoM description of the antenna
\begin{equation}
\Ivec = \Ymat \Vvec,
\label{eq:MoMdef}
\end{equation}
with~$\Ymat = \Zmat^{-1} \in \mathbb{C}^{N\times N}$ being admittance matrix, is related to port voltages~$\Vport \in \mathbb{C}^{P\times 1}$ via
\begin{equation}
\Vvec = \DIMmat \PORTmat \Vport,
\label{eq:Vportdef}
\end{equation}
where port positions are defined by an indexing matrix~$\PORTmat \in \left\{0,1\right\}^{N\times P}$,
\begin{equation}
C_{np} = \left\{ {\begin{array}{*{20}{cl}}
1 & \textrm{$p$-th port is placed at $n$-th position,} \\
0 & \textrm{otherwise},
\end{array}} \right.
\label{eq:PortMatdef}
\end{equation}
\ie{}, $\PORTmat^\herm\PORTmat = \M{1} \in \mathbb{R}^{P\times P}$. Since the basis functions~\eqref{eq:Zmatdef} may or may not have dimensions, diagonal normalization matrix~$\DIMmat \in \mathbb{R}^{N \times N}$ is defined elementwise as
\begin{equation}
D_{nn} = \xi_n
\label{eq:DimMatdef}
\end{equation}
to ensure that the port-based quantities such as impedances, voltages and currents have dimensions of Ohms, Volts and Amperes. For example, for dimensionless basis functions, such as RWG~\cite{RaoWiltonGlisson_ElectromagneticScatteringBySurfacesOfArbitraryShape}, the normalization variable would typically be the basis function's edge length~$l_n$, \ie{}, $\xi_n = l_n$. Analogously to the port voltage~$\Vport$, port current~$\Iport$ is defined via
\begin{equation}
\Iport = \PORTmat^\herm \DIMmat^\herm \Ivec.
\label{eq:Iportdef}
\end{equation}

Substituting~\eqref{eq:MoMdef} and, subsequently,~\eqref{eq:Vportdef} into~\eqref{eq:Iportdef} gives port admittance matrix~$ \Yport$
\begin{equation}
\Iport = \Yport \Vport,
\label{eq:YportDef}
\end{equation}
where
\begin{equation}
\Yport = \PORTmat^\herm \DIMmat^\herm \Ymat \DIMmat \PORTmat.
\label{eq:YportDef2}
\end{equation}
Note that complex power~\cite{Harrington_TimeHarmonicElmagField} is strictly conserved between port-like and MoM-like quantities, \ie{}, $\Iport^\herm \Vport = \Ivec^\herm \Vvec$.

The next step is the evaluation of the radiated power~\eqref{eq:Praddef2} using port voltages. This can be done thanks to the relation
\begin{equation}
\Ivec^\herm \M{M} \Ivec = \Vport^\herm \M{n} \Vport,
\label{eq:MportDef}
\end{equation}
with
\begin{equation}
\M{n} = \PORTmat^\herm \DIMmat^\herm \Ymat^\herm \M{M} \Ymat \DIMmat \PORTmat,
\label{eq:gDef}
\end{equation}
which is valid for any matrix~$\M{M}$ and results from substituting~\eqref{eq:MoMdef} and~\eqref{eq:Vportdef} into the left-hand side of~\eqref{eq:MportDef}. A particularly important example of this relation is the substitution~$\M{M} = \RmatVac$ which gives rise to relation
\begin{equation}
\Prad = \dfrac{1}{2} \Ivec^\herm \RmatVac \Ivec = \dfrac{1}{2} \Vport^\herm \Gport \Vport,
\label{eq:R0portDef}
\end{equation}
with~$\Gport =  \PORTmat^\herm \DIMmat^\herm \Ymat^\herm \RmatVac \Ymat \DIMmat \PORTmat$ and analogously for ohmic losses to
\begin{equation}
\Plost = \dfrac{1}{2} \Ivec^\herm \RmatMat \Ivec = \dfrac{1}{2} \Vport^\herm \Lport \Vport.
\label{eq:RmatportDef}
\end{equation}

The last step is the connection of power waves~$\apar, \bpar$, which exist in lossless feeding transmission lines of characteristic impedance~$R_{0,p}$, see Fig.~\ref{fig:fig1}, with port voltages, \ie{},
\begin{align}
\apar &= \dfrac{1}{2} \left( \ZcharMat^{-1} \Vport + \ZcharMat \Iport \right) = \dfrac{1}{2} \left(\M{1} + \ZcharMat \left( \Yport + \YportL \right) \ZcharMat \right) \ZcharMat^{-1} \Vport = \KportMat \Vport, \label{eq:adef} \\
\bpar &= \dfrac{1}{2} \left( \ZcharMat^{-1} \Vport - \ZcharMat \Iport \right) = \dfrac{1}{2} \left(\M{1} - \ZcharMat \left( \Yport + \YportL \right) \ZcharMat \right) \ZcharMat^{-1} \Vport = \LportMat \Vport, \label{eq:bdef}
\end{align}
where~\cite{Pozar_MicrowaveEngineering}
\begin{equation}
\Lambda_{pp} = \sqrt{R_{0,p}}
\label{eq:charImpdef}
\end{equation}
and where~admittances~$\J B_{\T{L},p}$ were accumulated at the diagonal of matrix~$\YportL$.

The final expression for TARC used throughout this paper reads
\begin{equation}
\TARC = \sqrt{1 - \dfrac{\Vport^\herm \Gport \Vport}{\Vport^\herm \KportMat^\herm \KportMat \Vport}} = \sqrt{1 - \dfrac{\apar^\herm \KportMat^{-\herm} \Gport \KportMat^{-1} \apar}{\apar^\herm \apar}}.
\label{eq:TARCdef2}
\end{equation}

The formulas~\eqref{eq:R0portDef},~\eqref{eq:RmatportDef}, and~\eqref{eq:TARCdef2} can, in principle, be evaluated in contemporary electromagnetic simulators as well. Nevertheless, the port-mode matrix formulation~\eqref{eq:gDef} seems to not be implemented yet. For this reason, the ohmic losses extraction is conventionally done via far field integration which is a time-consuming task. The procedure~\eqref{eq:gDef} is not only computationally more efficient but also more general. Any quantity based on matrix operator, \eg{}, stored energy matrix~\cite{CapekGustafssonSchab_MinimizationOfAntennaQualityFactor}, can be transformed. In this last case, only integration of near field, a computationally challenging operation, can circumvent the usage of~\eqref{eq:gDef}.

\subsection{Example -- Evaluation of TARC (Single-port Antenna)}

\begin{figure}
\centering
\includegraphics[width=\columnwidth]{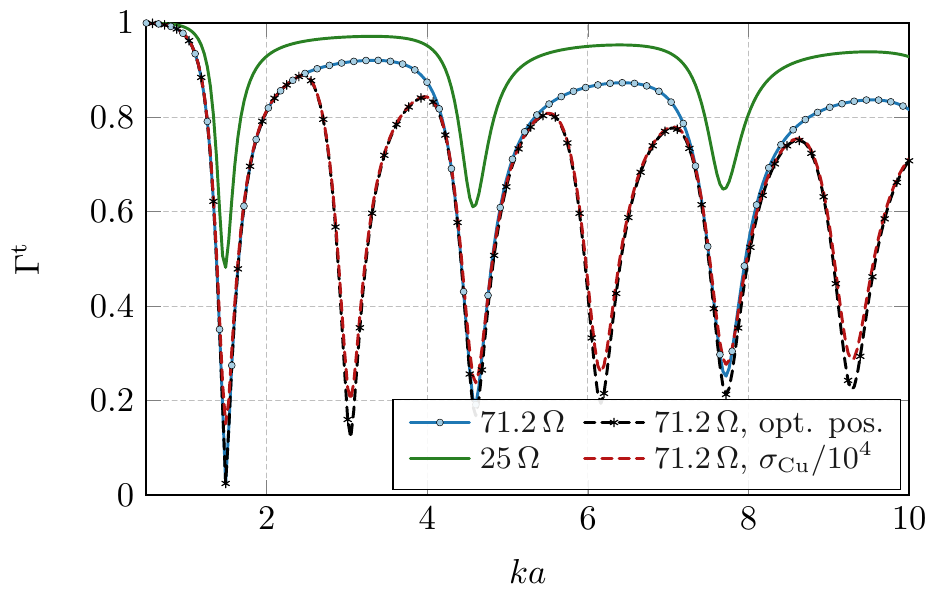}
\caption{TARC for a thin-strip dipole of length~$\ell$ and width~$\ell/100$. Various characteristic impedances~$R_{0,1}$ and surface resistivities were used. The dashed lines present the cases of the optimal delta gap placement at each frequency~$ka$. The solid curves assume the delta gap in the geometrical center of the dipole.}
\label{fig:dipoleTARC}
\end{figure}

\begin{figure}
\centering
\includegraphics[width=\columnwidth]{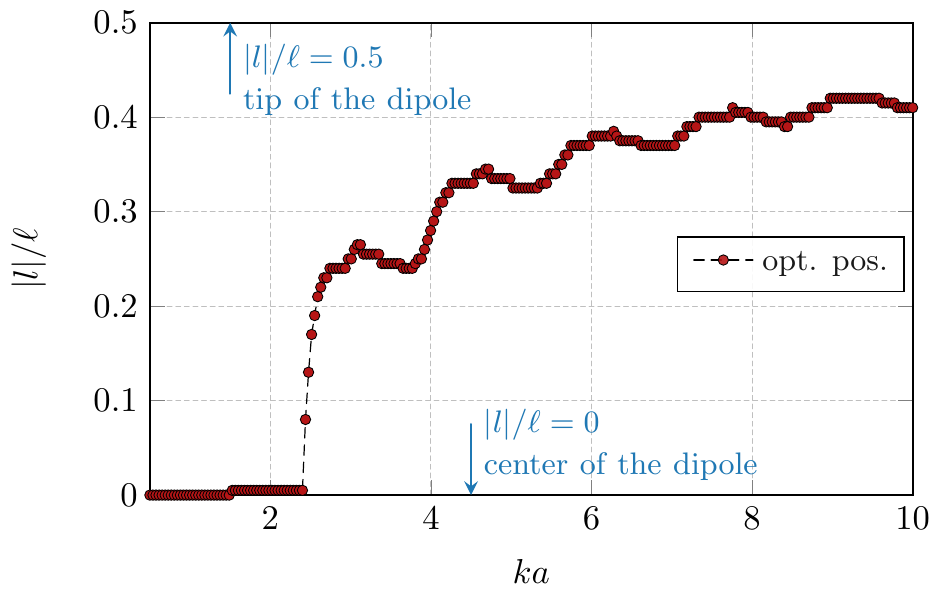}
\caption{Optimal placement of a delta-gap feeder along the thin-strip dipole made of copper. The placement is shown in relative length $|l|/\ell$ from the middle of the dipole. The vertical scale represents the position of the basis function to be fed at each electrical size~$ka$. The corresponding TARC is depicted by the black dashed line in Fig.~\ref{fig:dipoleTARC}.}
\label{fig:dipolePosition}
\end{figure}

Let us start with a simple single-port radiator, a thin-strip dipole made of copper ($\sigma_\T{Cu} = 5.96\cdot10^7\,\T{Sm}^{-1}$) with length~$\ell$ and width~$\ell/100$. The frequency range used for the study is~$ka \in [1/2, 10]$, where~$k$ is the wavenumber in vacuum and $a$~is a radius of the smallest sphere circumscribing the antenna. TARC is evaluated with the delta-gap feeding~\cite{Balanis1989} covering the entire width of the dipole. All the operators were evaluated in AToM~\cite{atom} which utilizes RWG basis functions~\cite{RaoWiltonGlisson_ElectromagneticScatteringBySurfacesOfArbitraryShape}. Two distinct values of transmission line characteristic impedance are studied,~$R_{0,1} = 71.2\,\Omega$ (the input impedance of the dipole at its first resonance) and $R_{0,1} = 25\,\Omega$, see Fig.~\ref{fig:dipoleTARC}. There are no tuning lumped elements, $\YportL = \M{0}$. Since the dipole is a single-port antenna, the specific excitation voltage plays no role in the evaluation of~\eqref{eq:TARCdef2} and can be freely set to~$v_1 = 1$\,V. This makes it possible to effectively find an optimal placement of the port along the dipole depending on the electrical size~$ka$. Matrices~$\Gport$ and~$\KportMat$, corresponding to a port at $n$-th edge, are denoted as~$g_{\srcRegion,n}$ and~$k_{\T{i},n}$ (they are scalars in this single-port case). Consequently, the minimal TARC, with respect to the optimal placement of the port, can be found via elementwise division
\begin{equation}
\min\limits_n \left\{ \Gamma_n^\T{t} \right\} = \max\limits_n \left\{ \dfrac{g_{\srcRegion,n}}{|k_{\T{i},n}|^2}\right\}
\label{eq:minimalTARCHadamard}
\end{equation}
for all tested positions. The position for minimum TARC is shown in Fig.~\ref{fig:dipolePosition} with the corresponding TARC shown in Fig.~\ref{fig:dipoleTARC}. 

The last study reveals that the dipole antenna is relatively immune to ohmic losses since the reduction of the conductivity by four orders in magnitude causes only a mild drop in performance. It is worth noting that, as far as the ohmic losses are negligible, TARC can be zeroed by a proper choice of tuning susceptance and characteristic resistance. This possibility is studied later on in Section~\ref{sec:terminImpend} for multi-port antennas.

\subsection{Example -- Evaluation of TARC (Multi-port Antenna)}
\label{sec:ex:TARC1}

\begin{figure}
\centering
\includegraphics[width=\columnwidth]{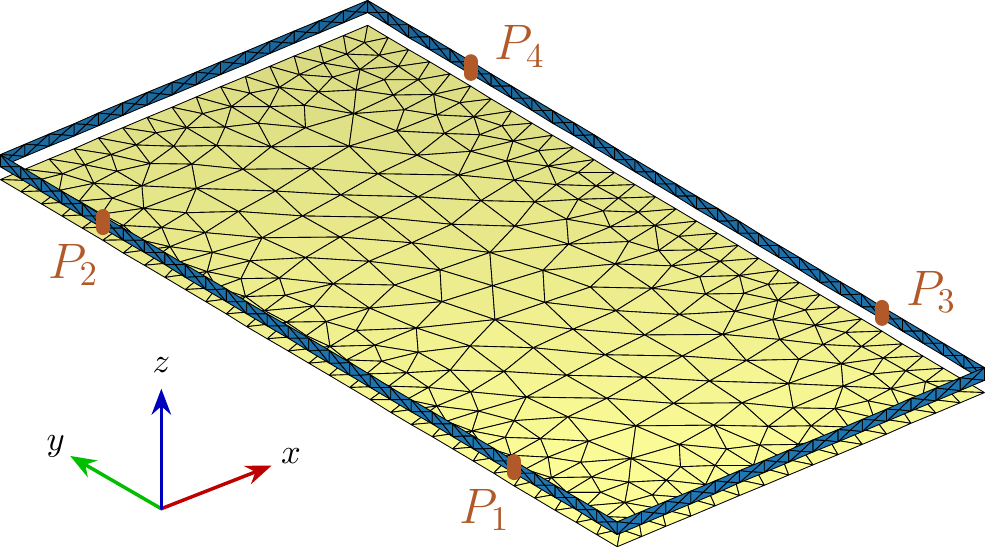}
\caption{A metallic rim with parasitic ground plane with four possible ports, denoted $P_p$, $p\in\left\{1,\dots,4\right\}$. Both the rim and the ground plane are made of copper. The ports are placed at a distance of~$\ell/5$ from the ends of the longer side.}
\label{fig:RIMstructure1}
\end{figure}

The second example, to be studied in the rest of the paper, is a four-port metallic rim placed above a ground plane, both of which are made of copper ($\sigma_\T{Cu} = 5.96\cdot10^7\,\T{Sm}^{-1}$), see Fig.~\ref{fig:RIMstructure1}. The dimensions of the structure are: length~$\ell = 150\,$mm, width~$\ell/2$, height of the strip~$3\ell/200$, and the elevation of the rim above the ground plane~$3\ell/200$. The rim is discretized by a uniform mesh grid consisting of $450$~basis functions. The ground plane is discretized by a Delaunay triangulation~\cite{deLoeraRambauSantos_Triangulations} with~$798$ basis functions, \ie{}, the total number of degrees of freedom is~$N=1248$. Notice here that the example serves mainly as a demonstration of a new designing framework for the effective evaluation and optimization of TARC. We have no intention of designing and optimizing a realistic system.

We start the investigation with the fixed placement of the ports, denoted as~$P_p$, $p\in\left\{1,\dots,4\right\}$, see Fig.~\ref{fig:RIMstructure1}, and with unit excitation~$v_p = 1\,\T{V}$ at all ports (the impressed electric field intensity points in $y$~direction for all ports) and with equal characteristic impedances~$R_{0,p}$.

\begin{figure}
\centering
\includegraphics[width=\columnwidth]{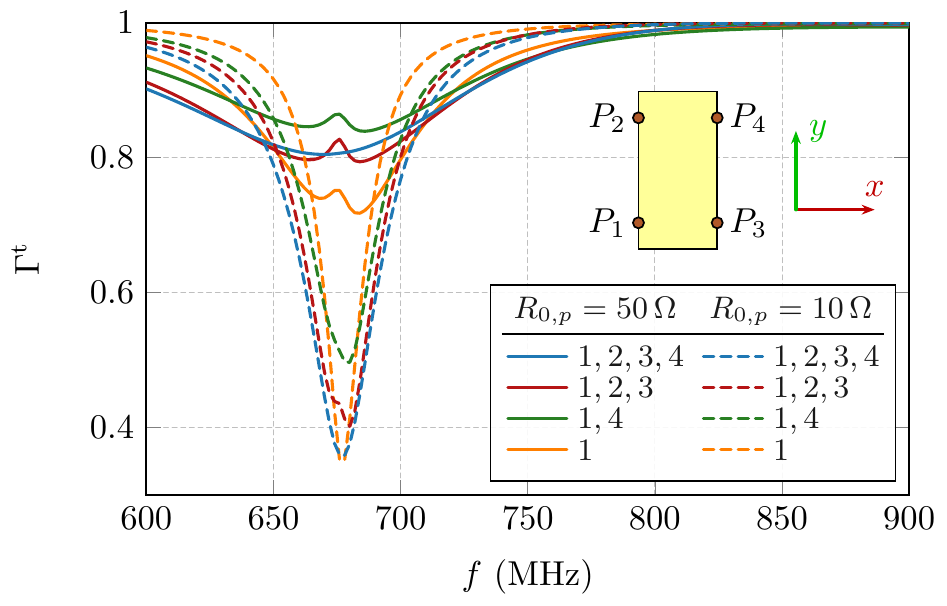}
\caption{TARC for the metallic rim with parasitic ground plane, depicted in Fig.~\ref{fig:RIMstructure1}. The characteristic impedance $50\,\Omega$ and $10\,\Omega$ were used for all enabled ports. Four various combination of ports with unit voltage excitation were studied.}
\label{fig:RIMstructureTARC1}
\end{figure}

TARC for two different impedances~$R_{0,p}$ and a varying number of uniformly excited ports is depicted in Fig.~\ref{fig:RIMstructureTARC1}. No lumped susceptances were used for simplicity, $B_{\T{L},p} = 0$. Some observations may already be made. In general, utilizing more ports does not automatically result in a lower value of TARC. Considering the fixed body of an antenna, optimal TARC is a complicated function of the characteristic impedances, matching, excitation, and ports' placement. The optimality of TARC with respect to all these parameters is studied in the following sections.

\section{Optimal Excitation for Minimum TARC}
\label{sec:TARCbound}
Let us first consider that characteristic impedances~$R_{0,p}$ and matching susceptances~$B_{\T{L},p}$ are fixed. In such scenario, TARC~\eqref{eq:TARCdef2} is, for a fixed geometry, solely a function of voltage vector~$\Vport$ and its minimization takes the form of the maximization of the total efficiency~$\eta_\T{tot}$ since
\begin{equation}
\eta_\T{tot} = \eta_\T{rad} \eta_\T{match} = 1 - \left(\TARC\right)^2 = \dfrac{\Vport^\herm \Gport \Vport}{\Vport^\herm \KportMat^\herm \KportMat \Vport},
\label{eq:totEffdef}
\end{equation}
where the radiation and matching efficiencies are defined as~\cite{Balanis_Wiley_2005}
\begin{equation}
\eta_\T{rad} = \dfrac{\Prad}{\Prad + \Plost}, 
\label{eq:radEffdef}
\end{equation}
and
\begin{equation}
\eta_\T{match} = \dfrac{\Prad + \Plost}{\Ptot}.
\label{eq:reflEffdef}
\end{equation}
Rigorously, the optimization problem for the maximal total efficiency (minimum TARC) reads
\begin{equation}
\begin{aligned}
	& \T{maximize} && \Vport^\herm \Gport \Vport \\
	& \T{subject\,\,to} && \Vport^\herm \KportMat^\herm \KportMat \Vport = 1,
\end{aligned}
\label{eq:totEffoptim}
\end{equation}
which is a quadratically constrained quadratic program~\cite{BoydVandenberghe_ConvexOptimization} solved 
by taking the largest eigenvalue~$\eta_1$
\begin{equation}
\max\limits_{\Vport} \left\{\eta_\T{tot}\right\} = \min\limits_{\Vport} \left\{ \TARC \right\} = \max_i \left\{ \eta_i \right\} = \eta_1
\label{eq:totEffBound2}
\end{equation}
of the generalized eigenvalue problem
\begin{equation}
\Gport \Vport_i = \eta_i \KportMat^\herm \KportMat \Vport_i.
\label{eq:totEffBound1}
\end{equation}
Eigenvector~$\Vport_1$, corresponding to the largest eigenvalue, represents the optimal terminal voltages. The optimal vector of incident power waves can be evaluated from the optimal port voltage as $\apar_1 = \KportMat \Vport_1$, see~\eqref{eq:adef}.

\begin{figure}
\centering
\includegraphics[width=\columnwidth]{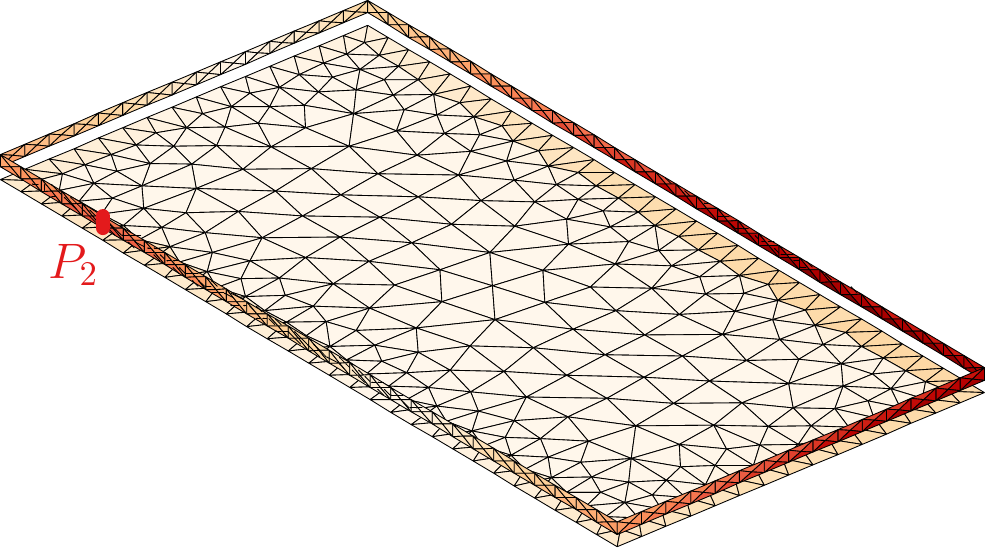}
\caption{Current density (absolute value) of a port-mode excited at $500\,$MHz by a unit voltage impressed at port $P_2$. For the set of ports depicted in Fig.~\ref{fig:RIMstructure1}, the excitation scheme reads~$\M{v} = \left[0\,\,1\,\,0\,\,0\right]^\herm$.}
\label{fig:RIMstructurePortMode}
\end{figure}

Optimal vector~$\Vport_1$ attains an interesting interpretation when relations~\eqref{eq:MoMdef}, \eqref{eq:Vportdef} are combined into
\begin{equation}
\Ivec = \Ymat \DIMmat \PORTmat \Vport = \sum\limits_p \Ivec_p v_p,
\label{eq:PortModedef}
\end{equation}
where~$\Ivec_p = \Ymat_p D_p$ denotes that the normalized column of matrix~$\Ymat$ connected to the $p$th port, the so-called \Quot{port mode} \cite{1978_Harrington_TAP}. Within the full-wave solution to the antenna problem, optimal port voltages~$\Vport_1$ thus excite a specific combination of port modes. An example of a port mode excited by port~$P_2$ is depicted in Fig.~\ref{fig:RIMstructurePortMode}. 

\subsection{Example -- Optimal Excitation of Multi-port Antenna}
\label{sec:ex:TARC2}

\begin{figure}
\centering
\includegraphics[width=\columnwidth]{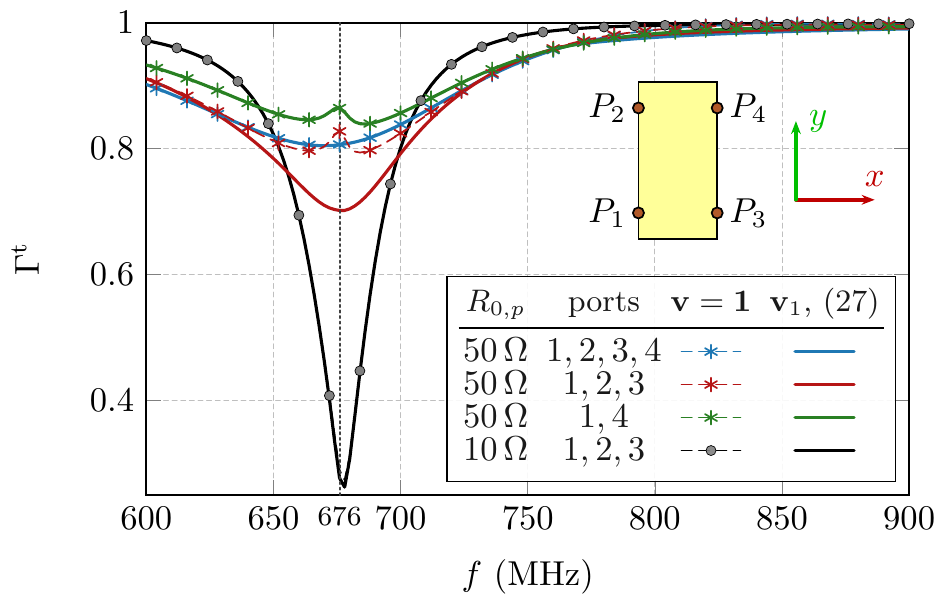}
\caption{Comparison of TARC for optimal excitation (solid lines) and uniform excitation (dashed lines). The characteristic impedance~$R_{0,p}=50\,\Omega$ was used for 4, 3, and 2 ports enabled, and $R_{0,p}=10\,\Omega$ for 3 ports. The configuration of ports is shown in the inset.}
\label{fig:RIMstructureTARC2}
\end{figure}

\begin{figure}
\centering
\includegraphics[width=\columnwidth]{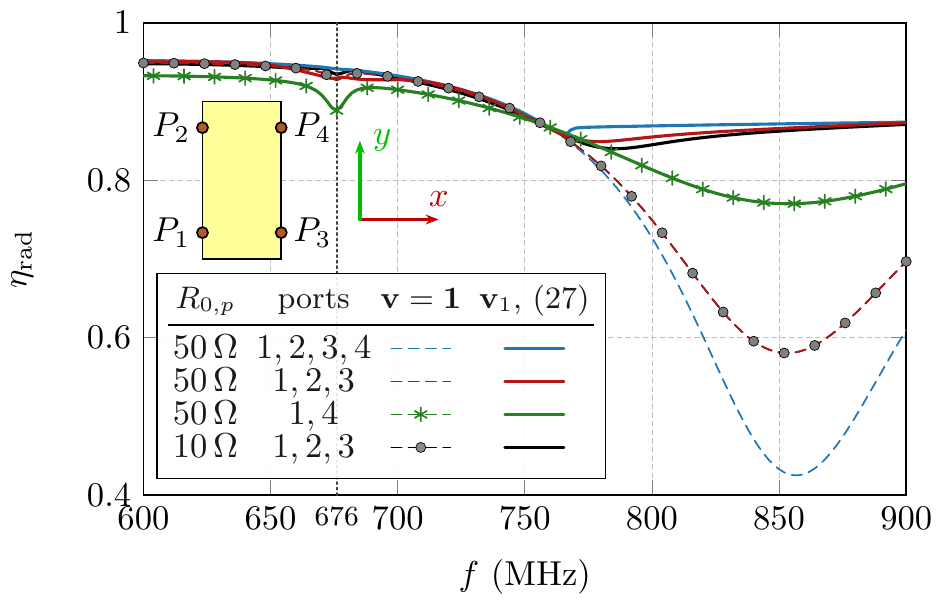}
\caption{Comparison of radiation efficiency for optimal excitation (solid lines) and uniform excitation (dashed lines). The setting is identical as in Fig.~\ref{fig:RIMstructureTARC2}.}
\label{fig:RIMstructureTARC3}
\end{figure}

Continuing with the example of the rim above a ground plane from Section~\ref{sec:ex:TARC1}, the optimal excitation voltages for 4, 3, and 2 ports with $R_{0,p}=50\,\Omega$ and 3 ports with $R_{0,p}=10\,\Omega$ are evaluated using~\eqref{eq:totEffBound1}, see Fig.~\ref{fig:RIMstructureTARC2}. The ports have the same configuration as in Fig.~\ref{fig:RIMstructureTARC1}. For the sake of convenience, corresponding results for unit voltages from Fig.~\ref{fig:RIMstructureTARC1} are directly included as dashed lines to evaluate the effect of optimal excitation. Radiation efficiency~$\eta_\T{rad}$, defined in~\eqref{eq:radEffdef}, is depicted in Fig.~\ref{fig:RIMstructureTARC3} for the same setup. The radiation efficiency for uniform excitation is depicted by dashed marked lines.

\begin{figure}
\centering
\includegraphics[width=\columnwidth]{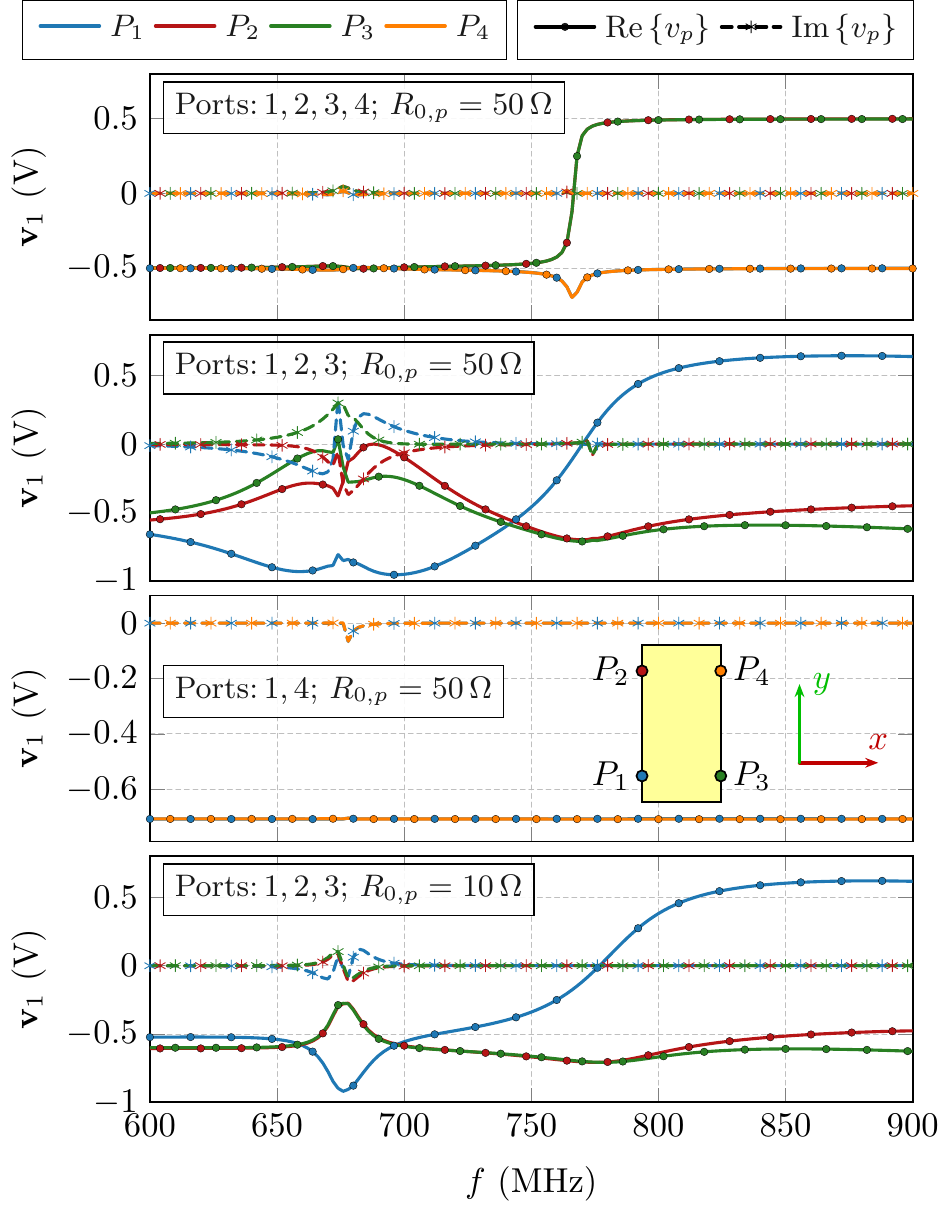}
\caption{Optimal excitation for configurations from Figs.~\ref{fig:RIMstructureTARC2} and \ref{fig:RIMstructureTARC3}.}
\label{fig:RIMstructureTARC5}
\end{figure}

It is seen that there is no improvement in TARC and radiation efficiency for four and two ports in the frequency range between~$600$\,MHz and~$770$\,MHz. This is a property of a specific placement of ports, which is very close to a point-symmetric configuration. If the symmetry was perfect, the solution to the eigenvalue problem~\eqref{eq:totEffBound1} would belong to one of the irreducible representations (irrep)~\cite{SchabBernhard_GroupTheoryForCMA}. The dominant solution in this frequency would belong to the irrep with unit voltages. The irrep is changed around~$770$\,MHz and non-uniform excitation found via~\eqref{eq:totEffBound1} improves the performance. This is most visible for radiation efficiency with 4 ports connected (blue lines, Fig.~\ref{fig:RIMstructureTARC3}). The corresponding optimal voltages and switch of irrep~\cite{Maseketal_ModalTrackingBasedOnGroupTheory} is depicted in the top pane of Fig.~\ref{fig:RIMstructureTARC5} which shows the optimal voltages for all the discussed scenarios.

\begin{figure}
\centering
\includegraphics[width=\columnwidth]{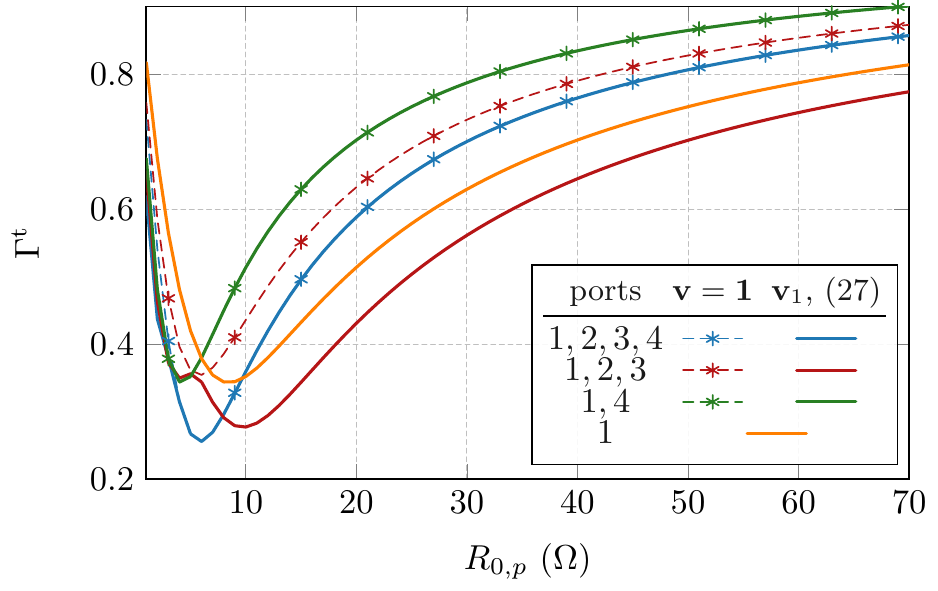}
\caption{Comparison of TARC for optimal excitation (solid lines) and uniform excitation (dashed marked lines) depending on the value of characteristic impedance~$R_{0,p}$. TARC is evaluated at frequency~$f=676\,\T{MHz}$.}
\label{fig:RIMstructureTARC4}
\end{figure}

Considering different characteristic impedances~$R_{0,p}$, the lower value $R_{0,p}=10\,\Omega$ leads to significantly better performance which is expected as the radiation resistance of the antenna is low. This observation addresses the question of optimal characteristic impedance. Figure~\ref{fig:RIMstructureTARC4} shows TARC for all four configuration of ports, both for unit (dashed) and optimal (solid) excitation depending on the value of characteristic impedance~$R_{0,p}$ for frequency~$f=676$\,MHz (highlighted by vertical dashed line in Figs.~\ref{fig:RIMstructureTARC2} and~\ref{fig:RIMstructureTARC3}). The best performance is found for four ports with $R_{0,p}\approx 5\,\Omega$ (blue line), \ie{}, for relatively low characteristic impedance. A slightly higher TARC is realizable with only three ports and $R_{0,p}\approx 10\,\Omega$ (red line). The latter option represents a practical and more appealing choice.

Thanks to the symmetrical arrangement for 2 and 4 port setups, the values of TARC for unit and optimal excitation coincide, which once again shows that the optimal excitation belongs to the irrep. with an in-phase constant eigen-vector, \ie{}, all delta gaps have the same amplitude at $f=676\,\T{MHz}$, see Fig.~\ref{fig:RIMstructureTARC5}, the top and second from the top panes. This observation underlines the need for further understanding of group theory and its role in solutions of integral equations~\cite{Maseketal_ModalTrackingBasedOnGroupTheory, SchabBernhard_GroupTheoryForCMA, Maseketal_ExcitationSchemesOfUncorrelatedChannels_Arxiv}.

It is seen that the fixed placement of ports specified by the designer significantly affects the overall performance of the antenna. Therefore, the next step is to investigate the optimal placement of the ports with simultaneous determination of their optimal excitation.

\section{Synthesis of Optimal Feeding Placement}
\label{sec:synthesis}

The maximal total efficiency (minimal TARC) given by the solution to~\eqref{eq:totEffBound1} is a function of number~$P$ and position~$\PORTmat$ of the ports, terminal impedances~$R_{0,p}$ and tuning reatances~$\J B_{\T{L},p}$. Considering a fixed number of ports~$P$, constant terminal impedance~$R_{0,p} = R_0$, and a given tuning susceptance~$\J B_{\T{L},p} = \J B_{\T{L}}$ for all ports, the only remaining variable is the optimal placement of the ports. This task is a feeding synthesis, defined as
\begin{equation}
\begin{aligned}
	& \T{maximize} && \eta_1 \\
	& \T{subject\,\,to} && \T{trace} \left( \PORTmat^\trans \PORTmat \right) = P, \\
\end{aligned}
\label{eq:feedSynthesisTARC}
\end{equation}
which is a combinatorial optimization problem solved by an exhaustive search (feasible only for small, though realistic, number of potential positions) or advanced optimization tools~\cite{Nemhauser_etal_IntegerAndCombinatorialOptimization}. The number of combinations is
\begin{equation}
\OP{C} \left(N,P\right) \equiv \binom{N}{P} = \dfrac{N!}{\left(N-P\right)! \, P!},
\label{eq:nCombin}
\end{equation}
where, in practice, $P$ is a small number, and $N$ being the number of potential positions to be tested. In order to truncate the solution space, only subregions with preferred positions of the ports might be specified (see the example below).

From the computational point of view, the matrices
\begin{align}
\label{eq:bigGmat}
\widehat{\M{G}}_\srcRegion &= \DIMmat^\herm \Ymat^\herm \RmatVac \Ymat \DIMmat \\
\label{eq:bigKmat}
\widehat{\M{K}}_\T{i} &= \dfrac{1}{2 \sqrt{R_0}} \left( \M{1} + R_0 \DIMmat^\herm \Ymat \DIMmat \right)
\end{align}
can be precalculated at the  beginning of the optimization procedure leaving only computationally cheap indexing operations
\begin{align}
\label{eq:smallgFromGmat}
\Gport &= \PORTmat^\herm \widehat{\M{G}}_\srcRegion \PORTmat \\
\label{eq:smallkiFromKmat}
\KportMat &= \PORTmat^\herm \widehat{\M{K}}_\T{i} \PORTmat
\end{align}
and dominant eigenvalue evaluation in~\eqref{eq:totEffBound1} to be performed for every combination.

\subsection{Example -- Optimal Placement of Feeding Ports}
\label{sec:optPlacement}

\begin{figure}
\centering
\includegraphics[width=\columnwidth]{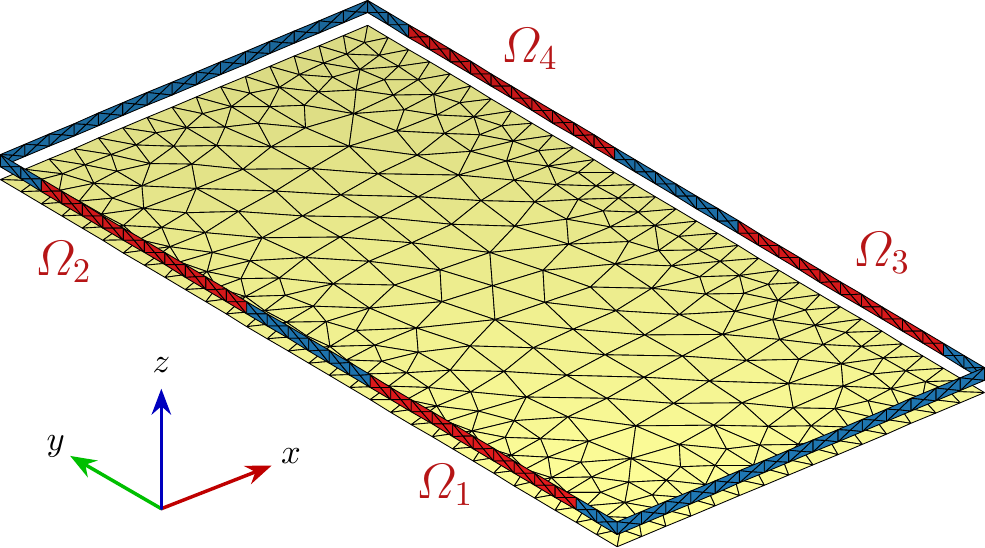}
\caption{The same metallic rim and parasitic ground plane as in Fig.~\ref{fig:RIMstructure1} with the regions for optimal ports' placement (red color). It is assumed that there is one or no port in each region, which always consists of 11 possible positions.}
\label{fig:RIMstructure2}
\end{figure}

The optimal placement of ports is investigated for a rim with the ground plane introduced in the previous sections. In order to imagine how complex the problem of the feeding synthesis is, assume first that all positions on the rim are suitable for accommodating delta gap feeding. The number of combinations is, in this case, $2^{90} \sim 1.24\cdot10^{27}$. Such an enormous number of combinations is reduced to $2^{44} \sim 1.76\cdot 10^{14}$ by assuming only subregions, as depicted in Fig.~\ref{fig:RIMstructure2}. The number of subregions for port placement is set to four, and it is assumed that in each subregion, denoted as $\srcRegion_i$, $i\in\left\{1,\dots,4\right\}$, contains not more than one port (no port is also an available option). With these restrictions, the number of solutions drops to $20735$. As a final reduction step, only unique arrangements are kept\footnote{There are combinations which represent the same arrangement of ports, only mirrored or rotated. This is a consequence of the symmetry of the rim, which belongs to the $C_{2\T{v}}$ group~\cite{McWeeny_GroupTheory}. These identical solutions can be found and truncated by application of symmetry operations which define a given group.}. This leads to only $5291$~combinations which have to be investigated.

\begin{figure}
\centering
\includegraphics[width=\columnwidth]{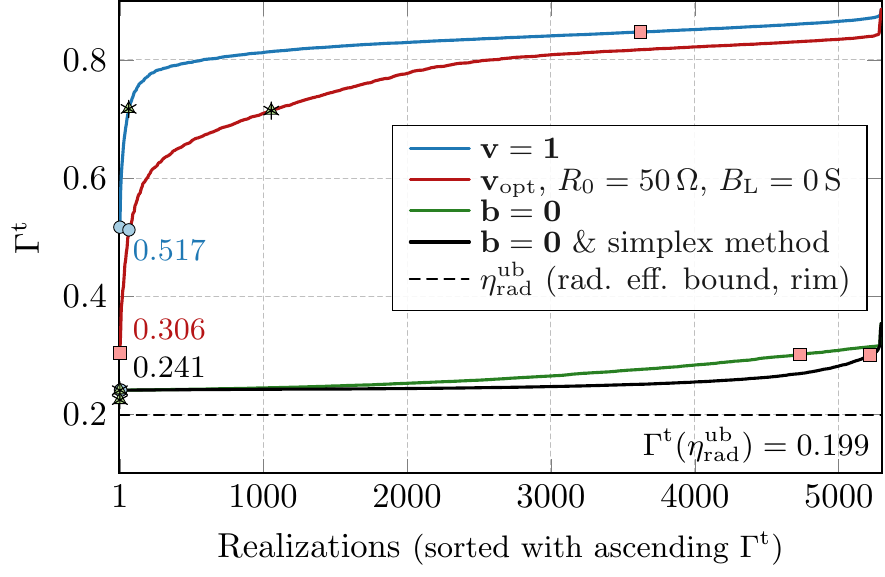}
\caption{Results of the feeding synthesis (solution to combinatorial optimization problem) for various approaches of TARC minimization. The meaning of all curves is explained in detail in the text of~Sec.~\ref{sec:optPlacement}. The best realization for each curve is shown in Fig.~\ref{fig:RIMplacement1} and highlighted by the marker. Other markers of the same shape show what is the performance of that combination of ports if another approach is utilized. For example, the best solution for the red curve performs relatively poorly for evaluations corresponding to the green and black curves. The dashed line indicates the fundamental bound on radiation efficiency (perfect matching is assumed), see Appendix~\ref{sec:bound}.}
\label{fig:RIMplacement2}
\end{figure}

\begin{figure}
\centering
\includegraphics[width=\columnwidth]{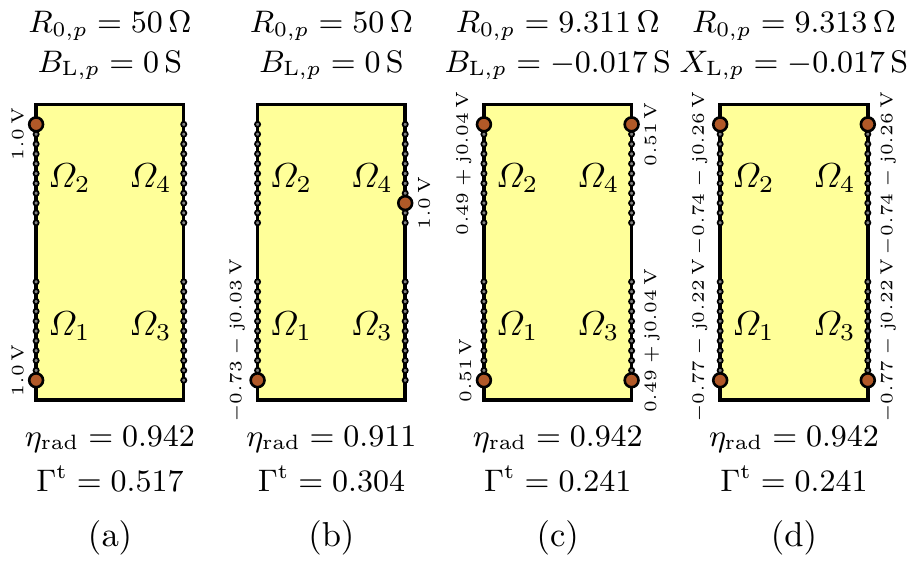}
\caption{An overview of the optimal placement of the ports and their performance in TARC and radiation efficiency. Various approaches found in this paper are utilized. The optimal placement defined in~\eqref{eq:feedSynthesisTARC} was found via an exhaustive search for all four cases: (a) uniform excitation, $\Vport=\M{1}$, (b) optimum excitation~$\Vport_1$ solved by~\eqref{eq:totEffBound1}, fixed value of $R_{0,p}$ and $B_{\T{L},p}$, (c) optimum excitation~$\Vport_i$, $R_{0,p}$, and $B_{\T{L},i}$ for $\bpar = \M{0}$ solved by \eqref{eq:bZero3}, (d) subsequent simplex optimization of~\eqref{eq:totEffBound1} after solving~\eqref{eq:bZero3}. The solutions (a)-(d) correspond to the best realizations in Fig.~\ref{fig:RIMplacement2} on the very left.}
\label{fig:RIMplacement1}
\end{figure}

\begin{figure}
\centering
\includegraphics[width=\columnwidth]{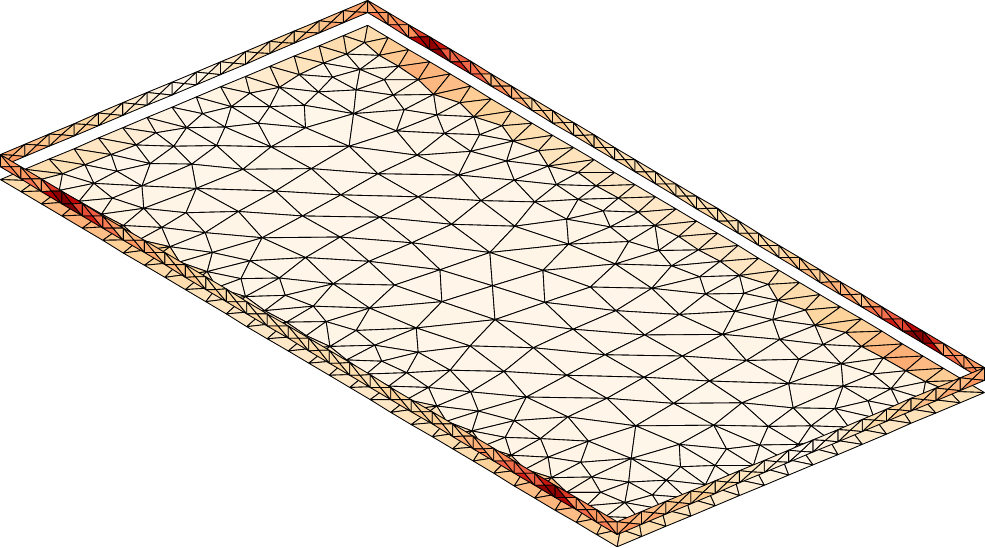}
\caption{Current density (absolute value) associated with the fundamental bound on radiation efficiency at frequency~$f=676$\,MHz for copper cladding, $\sigma=\sigma_\T{Cu}$, under the condition that only the current on the rim is controllable, see Appendix~\ref{sec:bound} for its evaluation. It is obvious that the placement of the ports from Fig.~\ref{fig:RIMplacement1}d correlates well with the maxima of the optimal current density.}
\label{fig:RIMfundBoundRadEff}
\end{figure}

Various approaches presented in this paper were applied to evaluate and optimize TARC for each aforementioned combination, see Fig.~\ref{fig:RIMplacement2}. The first approach is an application of unit voltages on all ports within the selected combination which is represented by the blue line in Fig.~\ref{fig:RIMplacement2}. It is seen that the overall performance in TARC is poor and the best combination, depicted in Fig.~\ref{fig:RIMplacement1}(a), reaches $\TARC = 0.517$. A modest improvement is realized by solving~\eqref{eq:totEffBound1} for~$R_{0,p}=50\,\Omega$ and~$B_{\T{L},p}=0\,\T{S}$ which results in the realization depicted in Fig.~\ref{fig:RIMplacement1}(b) reaching~$\TARC = 0.308$, see the red line in Fig.~\ref{fig:RIMplacement2}. The utilization of optimal feeding improves the performance, however, a further decrease in TARC is limited by the proper choice of circuit components. This is also indicated by an evaluation of the upper bound on radiation efficiency if the currents located on the rim are fully controllable and no tuning circuitry is involved, see Appendix~\ref{sec:bound}. The value of the upper bound at $f = 676\,\T{MHz}$ is $\eta_\T{rad}^\T{up} = 0.96$. Assuming perfect matching, $\eta_\T{match} = 1$, the corresponding TARC would be $\TARC = 0.199$, as indicated by the dashed black line in Fig.~\ref{fig:RIMplacement2}. The optimal current is depicted in Fig.~\ref{fig:RIMfundBoundRadEff} and the location of the current maxima justifies the choice for regions where the ports might be located, \cf{} Fig.~\ref{fig:RIMstructure2}. The next step, represented by the remaining curves, is to optimize the characteristic impedance of connected transmission lines and tuning susceptance.

\section{Optimal Characteristic and Tuning Impedances for Minimum TARC}
\label{sec:terminImpend}

For lossless antennas, the condition of perfect matching reads
\begin{equation}
\TARC = 0 \quad \Rightarrow \quad \bpar = \M{0}
\label{eq:bZero1}
\end{equation}
which, after substitution from~\eqref{eq:bdef}, gives
\begin{equation}
\left( \Yport + \YportL \right)\Vport = \left( \ZcharMat \ZcharMat \right)^{-1} \Vport.
\label{eq:bZero2}
\end{equation}
For a given excitation vector~$\Vport$ it is, therefore, always possible to find appropriate characteristic impedances and tuning susceptances to achieve perfect matching. Such solution might, however, lead to non-physical elements such as negative real part of characteristic impedance and furthermore assumes that all transmission lines are allowed to have different impedances. A more realistic scenario assumes that characteristic impedances of connected transmission lines are all identical~$\left( \ZcharMat \ZcharMat \right)^{-1} = R_\T{0L}^{-1} \M{1}$ and so are the tuning admittances~$\YportL =  \J B_\T{L} \M{1}$. Then the relation~\eqref{eq:bZero2} becomes an eigenvalue problem 
\begin{equation}
\Yport \Vport_i = \left( R_{\T{0L},i}^{-1} - \J B_{\T{L},i} \right) \Vport_i.
\label{eq:bZero3}
\end{equation}
with~$P$ solutions distinguished by index~$i$. The real part of the eigenvalues gives the reciprocal characteristic impedances (strictly positive), while the imaginary part gives the tuning susceptances which, together, constitute the generalization of the matching condition for a single-port antenna~\cite{Balanis_Wiley_2005}. It can be checked that for a single-port antenna the $1\times 1$~eigenvalue problem immediately gives~$B_{\T{L},1} = - B_\T{in}$ and $R_{0,1} = 1 / G_\T{in}$, where $Y_\T{in} = G_\T{in} + \J B_\T{in} = y$ is the input impedance of an antenna.

Considering no, or negligible, ohmic losses, combining~\eqref{eq:bZero2} with iteratively solved~\eqref{eq:feedSynthesisTARC} delivers optimal port placement, optimal voltage excitation, optimal sets of tuning susceptances and characteristic resistances. The only remaining task is to select, from all the available combinations of the ports' placement, the configuration which best fits the manufacturing and matching constraints. If none of the combinations is acceptable, the only other possibility is to change the shape of the antenna.

\subsection{Example -- Optimal Matching}
\label{sec:optMatching1}

The procedure from the previous section is utilized to further optimize TARC performance of the metallic rim. The formula~\eqref{eq:bZero3} is evaluated for all combinations of ports introduced in Section~\ref{sec:optPlacement}. For all solutions found, the optimal excitation and circuit parameters were used to evaluate true TARC value via~\eqref{eq:totEffdef}. The results are shown in Fig.~\ref{fig:RIMplacement2}, as illustrated by the green curve. A significant improvement is observed for all port combinations. The best solution offers~$\TARC = 0.241$, and its realization is shown in Fig.~\ref{fig:RIMplacement1}(c). Notice that TARC is not zeroed since the antenna is lossy and the assumption of zero loss from the previous section is not fulfilled. In a lossy case, the lowest TARC value is indicated by the fundamental bound on radiation efficiency in Fig.~\ref{fig:RIMplacement2}.

\begin{figure}
\centering
\includegraphics[]{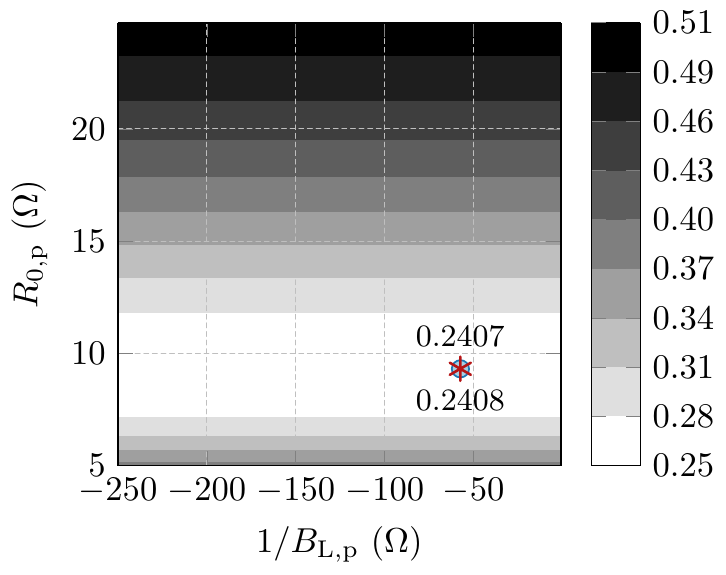}
\caption{TARC as a function of characteristic impedance~$R_{0}$ and tuning susceptance~$B_{\T{L}}$. The realization with the lowest TARC, \ie{}, the one depicted in Fig.\ref{fig:RIMplacement1}(d), is studied. The marks stand for the solution found by~\eqref{eq:bZero2} (circle mark, $\TARC=0.2408$) and for the solution found by~\eqref{eq:bZero2} with subsequent fine-tuning via simplex method~\cite{NocedalWright_NumericalOptimization} of~\eqref{eq:totEffBound1} (cross mark, $\TARC=0.2407$).}
\label{fig:RIMplacement2B}
\end{figure}

As a final step, each solution to~\eqref{eq:bZero3} was taken as an initial guess and entered the local optimization (simplex method) of~\eqref{eq:totEffBound1} with~$B_{\T{L}}$ and~$R_{0}$ being two real variables\footnote{In general, the optimized function is not convex in these variables and care must be taken to find globally optimal solution. In low-loss cases, the global optimum lies in the vicinity of perfectly matched setup and a local optimization is justified.}. This final optimum is, for each port combination, shown as the black curve in Fig.~\ref{fig:RIMplacement2}. Only mild improvement, as compared to the direct solution to~\eqref{eq:bZero3}, is observed. This indicates that the majority of the solutions were already very close to the minimum. This is confirmed by a parametric sweep of $B_{\T{L}}$ and $R_{0}$ for the globally best combination, represented by a star mark on the black curve in Fig.~\ref{fig:RIMplacement2} and by its realization in Fig.~\ref{fig:RIMplacement1}(d). The parameters, $B_{\T{L}}$ and $R_{0}$, were swept in a broad range around their optima and the optimal excitation with corresponding TARC were evaluated by~\eqref{eq:totEffBound1}. The values are shown in Fig.~\ref{fig:RIMplacement2B}, where the solution to~\eqref{eq:bZero3} and its further optimization via~\eqref{eq:totEffBound1} lie close to each other, \cf{}, Figs.~\ref{fig:RIMplacement1}(c) and Fig.~\ref{fig:RIMplacement1}(d). It can be seen from Fig.~\ref{fig:RIMplacement2B} that TARC, for this particular setting (structure, frequency, material, port placement), is relatively insensitive to variations in connected susceptance. Conversely, the precise realization of characteristic impedance is crucial to secure low TARC.

\begin{figure}
\centering
\includegraphics[width=\columnwidth]{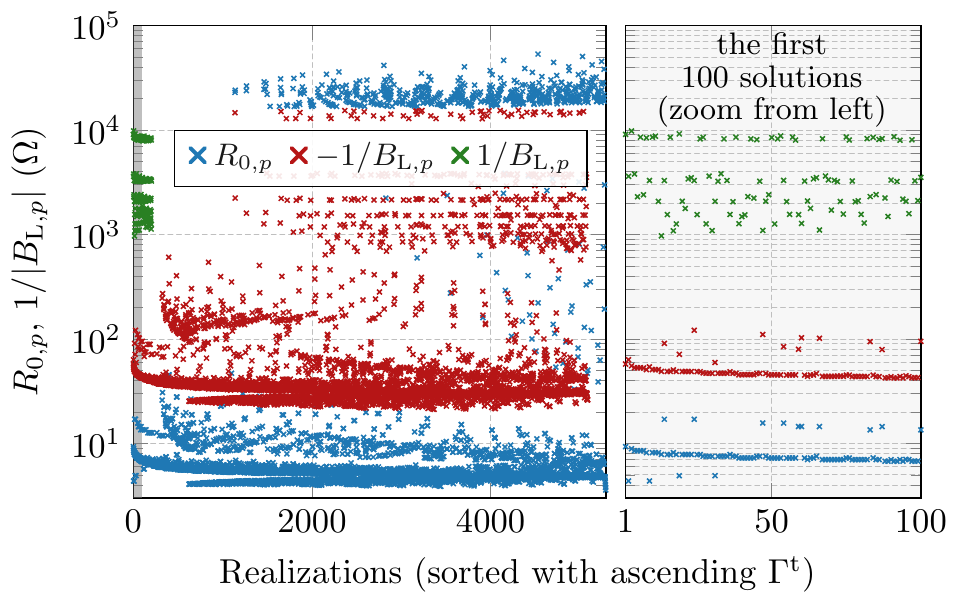}
\caption{The values of characteristic impedances and tuning elements required to reach TARC performance from Fig.~\ref{fig:RIMplacement2}, black curve. The realizations are arranged in the same order as realizations for the \Quot{$\bpar=\M{0}$, simplex method} curve in Fig.~\ref{fig:RIMplacement2}. A detail of the first 100 realizations with best performance in TARC are on the right. The elements with negative susceptance are highlighted by red cross marks, while those with positive susceptance are in green.}
\label{fig:RIMimpedances1}
\end{figure}

Since the particular choice of parameters~$B_{\T{L},p}$ and~$R_{0,p}$ is encumbered with some external constraints (manufacturing, availability of the components, etc.), the resulting optimal parameters of~$B_{\T{L},p}$ and~$R_{0,p}$ for all combinations of ports are shown in Fig.~\ref{fig:RIMimpedances1}. Their ordering is the same as for the data set in Fig.~\ref{fig:RIMplacement2} represented by the black curve, \ie{}, for the solution to~\eqref{eq:bZero3} with subsequent local optimization via~\eqref{eq:totEffBound1}. One can notice that positive susceptances (capacitors) are available only for a few solutions on the left. The characteristic impedances rarely overcome $20\,\Omega$ for this particular structure.

\begin{figure}
\centering
\includegraphics[width=\columnwidth]{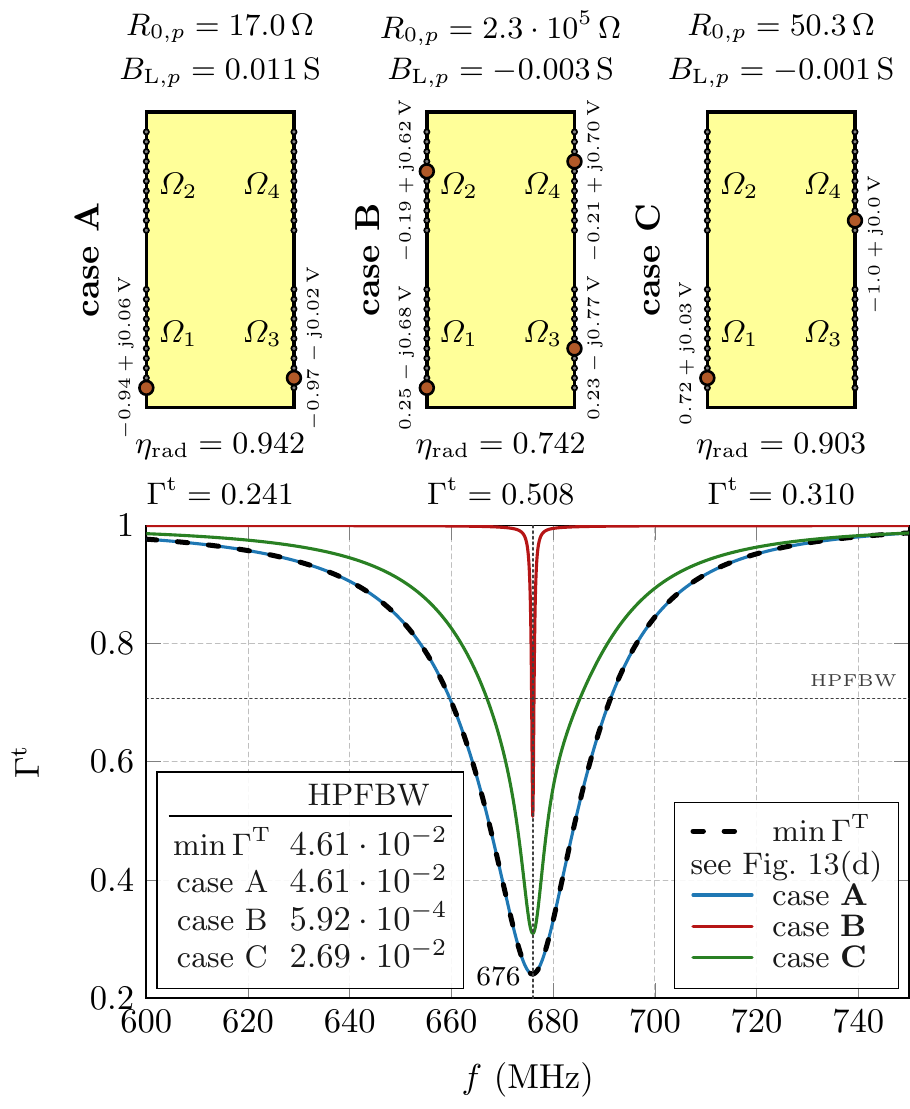}
\caption{Results of TARC~$\TARC$ for several realizations from Fig.~\ref{fig:RIMplacement2B}. The matching and excitation scheme for the minimal value of TARC reached at the frequency~$676$\,MHz (black dashed line) is depicted in Fig.~\ref{fig:RIMplacement1}(d). The matching and excitation schemes for cases~A, B, and C are depicted in top row of the figure. The case~A was chosen because of its TARC value is almost the same as for the optimal solution and the value of the resistance~$R_{0,p}$ is approximately doubled. The case~B was chosen because of its extremely high value of resistance~$R_{0,p}$. Finally, the case~C was chosen because of the closeness of its resistance~$R_{0,p} = 50.3\,\Omega$ to $50\,\Omega$. The level of half-power fractional bandwidth (HPFBW) is highlighted by the horizontal dotted black line with the values shown in the inset of the figure.}
\label{fig:RIMVariousMatching}
\end{figure}

Figures~\ref{fig:RIMplacement2} and \ref{fig:RIMimpedances1} suggest that there is a great variety of port placements reaching almost the same TARC value. Particularly, the first~$3000$ solutions from Fig.~\ref{fig:RIMplacement2} differ in TARC value by less than $2.5$\,\%. Their corresponding matching parameters shown in~Fig.~\ref{fig:RIMimpedances1}, however, differ considerably (ordering in both figures is the same) and with them the TARC bandwidth, see Fig.~\ref{fig:RIMVariousMatching}, which can be a parameter of the final choice.

\begin{figure}
\centering
\includegraphics[width=\columnwidth]{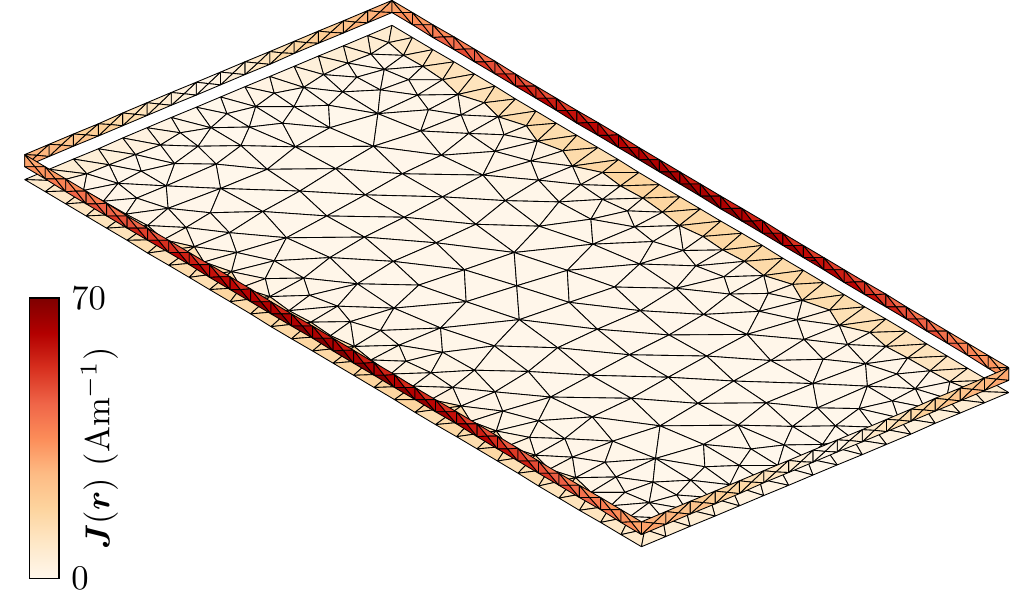}
\caption{Current density (absolute value) induced by the feeding scheme from Fig.~\ref{fig:RIMplacement1}(d), $f=676\,$MHz.}
\label{fig:RIMoptCurrentDens}
\end{figure}

\begin{figure}
\centering
\includegraphics[width=6cm]{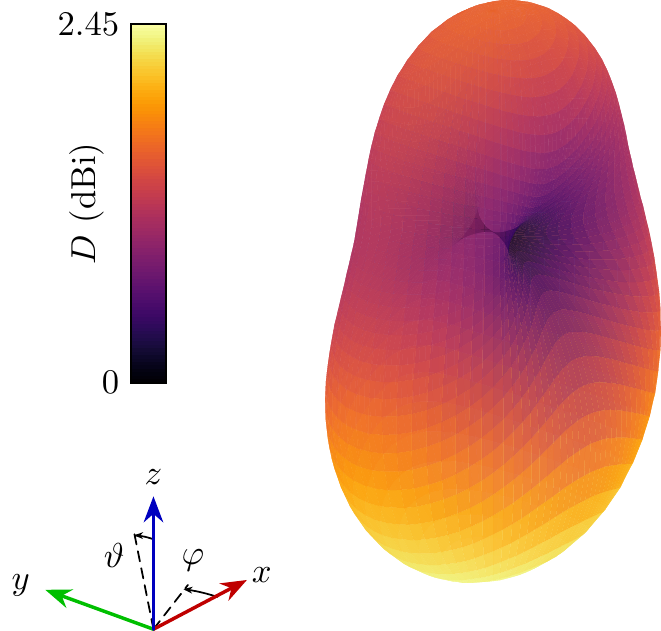}
\caption{Radiation pattern generated by a current from Fig.~\ref{fig:RIMoptCurrentDens}. The highest directivity in the broadside direction is~$D = 2.37\,$dBi.}
\label{fig:RIMoptFarField}
\end{figure}

To confirm the superb performance of the solution from Fig.~\ref{fig:RIMplacement1}(d), the fundamental bound on radiation efficiency was calculated for this particular combination of ports. The value of the upper bound~$\eta_\T{rad}^\T{up} = 0.942$ is equal to the radiation efficiency realized by the ports and circuitry from Fig.~\ref{fig:RIMplacement1}(d), \ie{}, there is no reflected power thanks to the solution to~\eqref{eq:bZero3} and the radiation efficiency reaches the fundamental bound. The current density and radiation pattern for the best solution is depicted in Fig.~\ref{fig:RIMoptCurrentDens} and Fig.~\ref{fig:RIMoptFarField}, respectively. In other words, there is no room for further improvement of the performance other than changing the shape of the antenna or adding additional ports.

This example demonstrated that there is a simple and straightforward technique to find an optimal placement of ports, their optimal excitation, and optimal parameters of a matching circuit which, for a fixed antenna body, prescribed losses and frequency, gives the lowest possible value of TARC.

\section{Realized Gain}
\label{sec:realGain}

Antenna directivity~\cite{Balanis_Wiley_2005} is defined here as~\cite{JelinekCapek_OptimalCurrentsOnArbitrarilyShapedSurfaces}
\begin{equation}
D\left(\UV{e},\UV{r}\right) = \dfrac{4 \pi}{Z_0} \dfrac{\Ivec^\herm \Fvec^\herm\Fvec \Ivec}{\Ivec^\herm \RmatVac \Ivec} = \dfrac{4 \pi}{Z_0} \dfrac{\Vport^\herm \Fport^\herm\Fport \Vport}{\Vport^\herm \Gport \Vport},
\label{eq:directdef}
\end{equation}
where~$F\left(\UV{e},\UV{r}\right) = \Fvec \left(\UV{e},\UV{r}\right) \Ivec$ is the electric far field in polarization~$\UV{e}$ and direction~$\UV{r}$,~$Z_0$ is the impedance of free space, and $\Fport = \Fvec \left(\UV{e},\UV{r}\right) \Ymat \DIMmat \PORTmat$ is port representation of far-field, see~\cite[App.~D]{JelinekCapek_OptimalCurrentsOnArbitrarilyShapedSurfaces} for details. Putting TARC~$\TARC$ and antenna directivity~$D$ together gives a realized gain (sometimes called absolute gain~\cite{Balanis_Wiley_2005}):
\begin{equation}
\Grealized = \left(1 - \left(\TARC\right)^2\right) D = \dfrac{4 \pi}{Z_0} \dfrac{\Vport^\herm \Fport^\herm\Fport \Vport}{\Vport^\herm \KportMat^\herm \KportMat \Vport} = \dfrac{4 \pi}{Z_0} \dfrac{\left|\Fport \Vport\right|^2}{\left|\KportMat \Vport\right|^2}.
\label{eq:realGain}
\end{equation}
The optimization problem for maximal realized gain reads
\begin{equation}
\begin{aligned}
	& \T{maximize} && \Vport^\herm \Fport^\herm \Fport \Vport \\
	& \T{subject\,\,to} && \Vport^\herm \KportMat^\herm \KportMat \Vport = \dfrac{Z_0}{4 \pi},
\end{aligned}
\label{eq:RGoptim}
\end{equation}
and is solved by a generalized eigenvalue problem
\begin{equation}
\Fport^\herm \Fport \Vport_i = \gamma_i \dfrac{Z_0}{4\pi} \KportMat^\herm \KportMat \Vport_i.
\label{eq:RGbound1}
\end{equation}
Since the LHS of~\eqref{eq:RGbound1} contains a rank-1 operator, the dominant eigenvector is known analytically as
\begin{equation}
\Vport_1 \propto \left(\KportMat^\herm \KportMat \right)^{-1} \Fport^\herm
\label{eq:RGbound2}
\end{equation}
and which, when substituted into the Rayleigh quotient of~\eqref{eq:RGbound1}, gives a maximal realized gain
\begin{equation}
\Grealized_\T{up} = \gamma_1 = \dfrac{4\pi}{\ZVAC} \left|\Fport \KportMat^{-1} \right|^2.
\label{eq:RGbound3}
\end{equation}
The maximal realized gain is a function of ports' placement, tuning susceptances, and characteristic impedances (through the matrix~$\KportMat$). For additional optimization of ports' placement, see Section~\ref{sec:synthesis} with~$\eta_1$ being changed to~$\gamma_1$ in~\eqref{eq:totEffBound1}. From the computational point of view, the matrix inversion has to be iteratively solved in \eqref{eq:RGbound3} instead of the determination of the dominant eigen-pair in~\eqref{eq:totEffBound2} or \eqref{eq:bZero2}.

\subsection{Example -- Optimal Excitation for Maximal Realized Gain}
\label{sec:realGainEX}

The four-element dipole antenna array operating at $f = 1\,\T{GHz}$ was chosen as a simple and instructive example. The thin-strip dipoles are of resonant length ($\ell = \lambda/2$), their width is $\ell/100$, and they are made of copper ($\sigma_\T{Cu} = 5.96\cdot 10^7\,\T{Sm}^{-1}$). Each dipole has a delta gap feeding in its center. Two arrangements are studied:
\begin{enumerate}
\item an uniform array with separation distance~$d = \lambda/2$,
\item a non-uniform array,~$d = \left\{\lambda/20, \lambda/10, \lambda/4\right\}$.
\end{enumerate}
Dipoles are, in both cases, parallel to the $z$-axis, the delta gap feeders are placed in a $z=0$ plane, and both arrays are centered at the origin of the coordinate system. The dominant polarization~$\UV{e} = \UV{\vartheta}$ is investigated in the~$\vartheta=\pi/2$, $\varphi=[0,\pi]$ half-cut, \ie{}, from the end-fire direction ($\varphi=0$, to $+x$~direction), through the broad-side direction ($\varphi=0$, to $+y$~direction), and ending with the end-fire direction ($\varphi=\pi$, to $-x$~direction).

\begin{figure}
\centering
\includegraphics[width=\columnwidth]{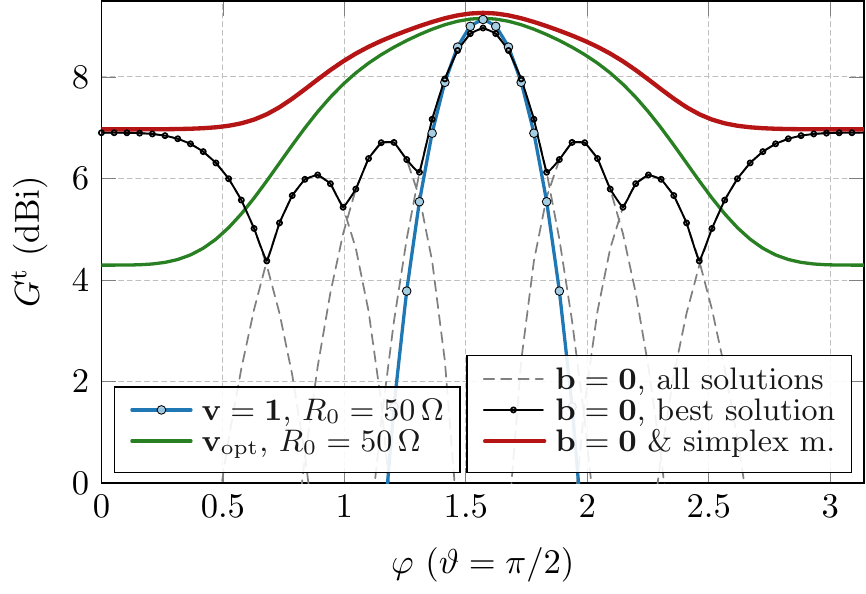}
\caption{Results of the feeding synthesis (solution to the optimization problem for angle~$\varphi$) for various approaches of realized gain maximization. The uniform array of four metallic dipoles, separated by distance~$d = \lambda/2$ is considered. The meaning of all curves is explained in details in Sec.~\ref{sec:realGainEX}. The blue line stands for unit feeding, $\M{v}=\M{1}$. The curve denoted as $\M{v}_\T{opt}$ (green solid) was evaluated by~\eqref{eq:RGbound3}. The black curves are individual solutions to~\eqref{eq:bZero3} (thin dashed) with their maximum envelope highlighted by the solid black line. The final solution is represented by the solid red line, which was found by solving~\eqref{eq:bZero3} with subsequent simplex optimization of~\eqref{eq:RGbound3}.}
\label{fig:arrayUniformMaxGain}
\end{figure}

\begin{figure}
\centering
\includegraphics[width=\columnwidth]{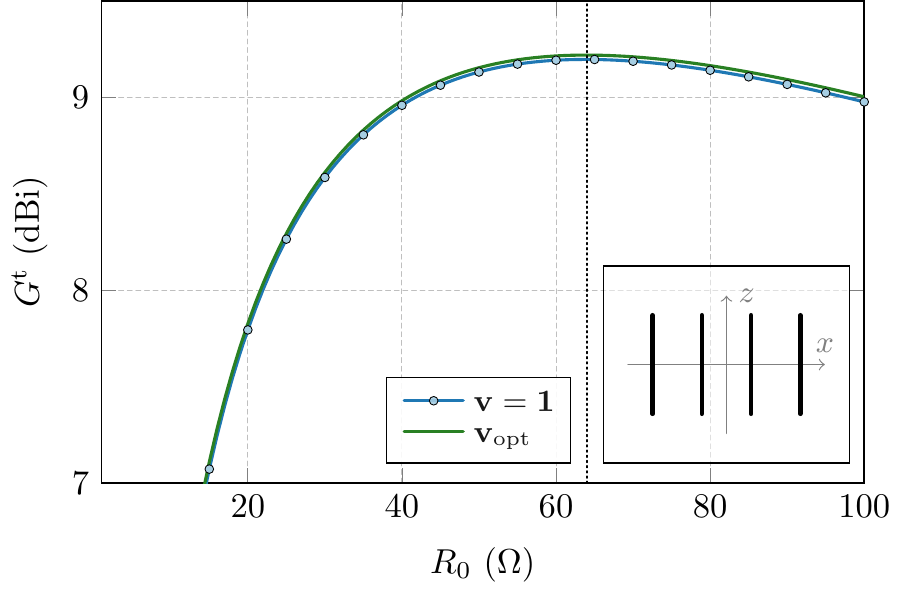}
\caption{Sensitivity of the realized gain~$G^\T{t}$ in the broad-side direction ($\varphi=\pi/2$, $\vartheta=\pi/2$) to the value of characteristic impedance~$R_0$. The meaning of the curves is the same as in the Fig.~\ref{fig:arrayUniformMaxGain}. The inset shows the uniform array studied in this example.}
\label{fig:arrayNonuniformMaxGainR0}
\end{figure}

The uniform array is treated first. The realized gain~$G^\T{t}$ is depicted in Fig.~\ref{fig:arrayUniformMaxGain} for various excitation schemes. The first scheme uses unit amplitudes at all ports which is favorable in the broad-side direction, but performs poorly for other directions~\cite{Balanis_Wiley_2005}. When the excitation is found by~\eqref{eq:RGbound3} (tuning circuitry fixed at $R_0 = 50\,\Omega$ and $B_\T{L} = 0\,\T{S}$) for each direction~$\varphi$, performance is improved for all studied directions except for the broad-side direction, where the unit excitation is already optimal. Still, relatively poor performance is observed in the end-fire direction, the reason being the fixed matching circuitry. The dependence on the characteristic impedance~$R_0$ is shown in~Fig.~\ref{fig:arrayNonuniformMaxGainR0} for a broad-side direction where it is seen that performance close to the optimum occurs in the relatively broad range of~$R_0$, with the maximum being reached around~$R_0 \approx 64\,\Omega$ (highlighted by the vertical dashed line). Major improvements are thus expected from the reactive matching via elements~$B_\T{L}$. This is accomplished by solving~\eqref{eq:bZero3} with a subsequent simplex optimization of~\eqref{eq:RGbound3}.

\begin{figure}
\centering
\includegraphics[width=\columnwidth]{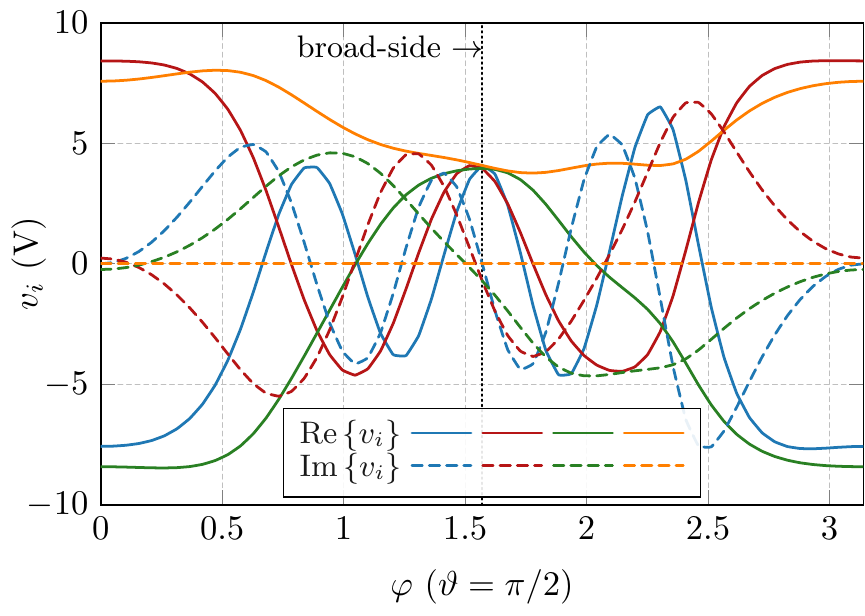}
\caption{The optimal excitation of the uniform metallic dipole array with separation distance~$d=\lambda/2$. Different colors represent real (solid) and imaginary (dashed) parts of the voltage~$v_i$ impressed at the $i$-th port. Optimal excitation with optimal circuit parameters~$R_0$ and~$B_\T{L}$ (not shown) was found for each direction~$\varphi$ via the solution to~\eqref{eq:bZero3} and the subsequent simplex optimization of~\eqref{eq:RGbound3}. Their realization led to the optimal performance indicated by the red curve in Fig.~\ref{fig:arrayUniformMaxGain}.}
\label{fig:arrayNonuniformMaxGainVopt}
\end{figure}

The four solutions to~\eqref{eq:bZero3} are depicted by black dashed lines with the maximum envelope highlighted by the black solid line. All these solutions were subsequently considered as the starting point for a simplex optimization method which tried to locally maximize~\eqref{eq:RGbound3}. Similarly as in the previous section, this technique is capable of delivering excellent results, represented here by the red solid line, which dominates the performance in realized gain~$G^\T{t}$ for all investigated directions. The associated optimal excitation of all four ports is shown in Fig.~\ref{fig:arrayNonuniformMaxGainVopt}. Notice the in-phased constant voltages for the broad-side direction and, conversely, the alternating polarity of the voltages for the end-fire direction (\ie, $v_i \sim \T{exp}\left\{-\J \pi i \right\}$, $i\in\left\{1,\dots,4\right\}$). The end-fire setup closely follows the phase progression opposite the electric distance between the radiators~($kd = \pi$) as recommended by textbooks~\cite{Balanis_Wiley_2005}.

\begin{figure}
\centering
\includegraphics[width=\columnwidth]{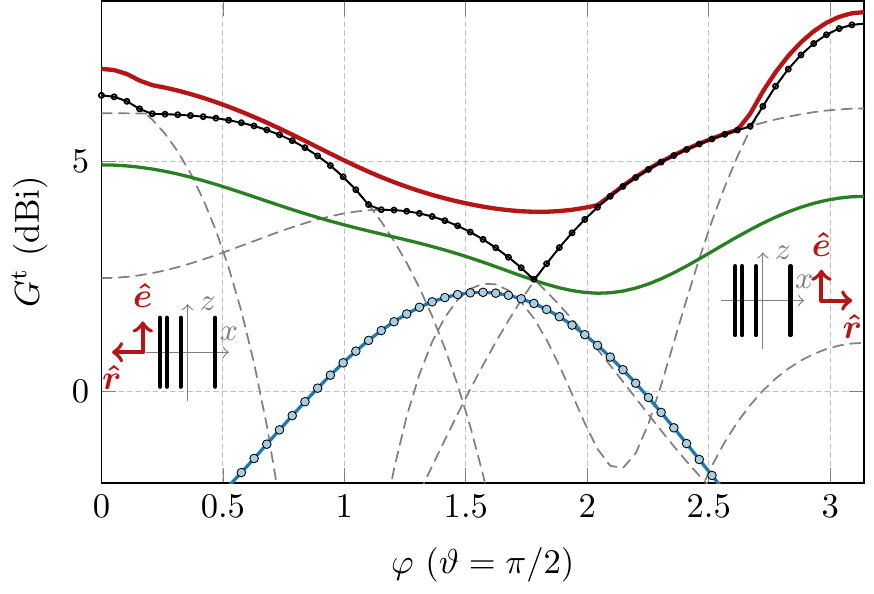}
\caption{Results of the feeding synthesis (solution to the optimization problem for angle~$\varphi$) for various approaches of realized gain minimization. The nonuniform array of four metallic dipoles, separated by distance~$d = \left\{\lambda/20,\lambda/10,\lambda/4\right\}$ is considered. The meaning of all curves is the same as in Fig.~\ref{fig:arrayUniformMaxGain}.}
\label{fig:arrayNonuniformMaxGain}
\end{figure}

As anticipated from the array theory~\cite{Elliot_AntennaTheoryAndDesign}, the ability to radiate well to a given direction changes with separation distance~$d$ between individual array elements. This is the case of the second array considered, for which the elements are nonuniformly spaced, separated by relatively short distance. The performance of such an array is shown in Fig.~\ref{fig:arrayNonuniformMaxGain} with the same meaning of all curves as in Fig.~\ref{fig:arrayUniformMaxGain}. The highest realized gain~$G^\T{t}$ is reached in end-fire directions, preferably in the~$+x$ direction. The best solution to~\eqref{eq:bZero3} is, in this case, very close to the solution refined by subsequent simplex optimization.

The two examples of thin-wire metallic arrays demonstrated effectiveness and insight gain by the utilization of the port mode-based definition and optimization of realized gain. Notice that the same analysis can be done for an arbitrarily large array or multi-port antenna possibly made of inhomogeneous and lossy materials.

\section{Conclusion}
\label{sec:concl}

The total active reflection coefficient (TARC) and realized gain were reformulated in terms of source current density and terminal voltages. This enabled the evaluation of these antenna metrics in full-wave fashion, taking into account ohmic losses of realistic metals and the particular shape(s) of radiator(s). There are no restrictions on homogeneity of materials. Since both figure of merits are defined in terms of quadratic forms, the presented formulation allows not only for the easy evaluation of TARC and realized gain, but also for the determination of optimal voltages, optimal placement of the ports, or determination of optimal terminal impedances and optimal matching.

Within this formulation, the full method of moments solution has to be evaluated only once and, while the admittance matrix is obtained, only computationally cheap indexation has to be performed to sweep the position of the ports. All resulting eigenvalue problems contain only small matrices and, in the case of realized gain, the optimal excitation for its maximization was even found analytically, being a sole function of ports' placement and characteristic and matching impedances.

The fast evaluation and subsequent optimization makes it possible to solve the combinatorial feeding synthesis problem for all practically reasonable configurations of ports with an exhaustive search. As a side-product, the optimal excitation and optimal circuitry is delivered. The designer then has full access to all possible solutions, their performance, and a list of all circuit elements needed. When performance is not sufficient, the only resolution required is a change of supporting geometry or the addition of extra ports.

The method was demonstrated on a simple dipole, a multi-port MIMO antenna, and uniform and nonuniform arrays. The presented results confirm the usefulness of the method for multi-port antenna systems with the advantage of having the possibility of including/excluding ohmic losses from the evaluation of TARC to make the results comparable with practical measurements.

Further research is needed to study other multi-port metrics as a cross-correlation coefficient or the ability to excite multiple states simultaneously and to define and optimize these quantities in the same way as shown in this paper. An interesting extension would be to combine this work with a selective excitation of orthogonal radiation patterns and to determine conditions for optimal performance. Establishing a connection to characteristic port modes can reveal a link between their selective excitation and optimality in TARC and realized gain. Another research direction is optimality within the frequency band. Finally, since both this work and the topology sensitivity approach~\cite{Capeketal_ShapeSynthesisBasedOnTopologySensitivity} shares the admittance matrix as the only variable needed to perform either feeding of topology synthesis, the possibilities of how to merge these two treatments of important antenna problems will be studied as an ultimate goal of antenna synthesis.

\appendices

\section{Fundamental Bound on Radiation Efficiency}
\label{sec:bound}

The fundamental bound on radiation efficiency~\eqref{eq:radEffdef} is expressed here in terms of the port modes associated with a controllable~\cite{CapekGustafssonSchab_MinimizationOfAntennaQualityFactor} subregion~$\srcRegion_\T{C} \subseteq \srcRegion$, represented by a particular indexing matrix~$\PORTmat$.

The procedure starts with the introduction of the dissipation factor~$\delta_\T{rad}$~\cite{Harrington_EffectsOfAntennaSizeOnGainBWandEfficiency} 
\begin{equation}
\eta_\T{rad} = \dfrac{1}{1 + \delta_\T{rad}},
\label{eq:radEffoptim1}
\end{equation}
which has a favorable scaling~$\delta_\T{rad} \in \left[0,\infty\right]$, for details see~\cite{GustafssonCapekSchab_TradeOffBetweenAntennaEfficiencyAndQfactor}. In order to accommodate the requirement on a controllable subregion~$\srcRegion_\T{C}$, the procedure from~\cite{JelinekCapek_OptimalCurrentsOnArbitrarilyShapedSurfaces} is applied to a port mode representation of the radiated and lost power, \eqref{eq:R0portDef} and \eqref{eq:RmatportDef}, as
\begin{equation}
\begin{aligned}
	& \T{minimize} && \Vport^\herm \Lport \Vport \\
	& \T{subject\,\,to} && \Vport^\herm \Gport \Vport = 1,
\end{aligned}
\label{eq:radEffoptim2}
\end{equation}
which is readily solved by an eigenvalue problem
\begin{equation}
\Lport \Vport_i = \delta_i \Gport \Vport_i.
\label{eq:radEffoptim3}
\end{equation}
The unknowns~$\Vport$ in \eqref{eq:radEffoptim2} and \eqref{eq:radEffoptim3} are the expansion coefficients of the ports modes~\eqref{eq:PortModedef} associated with a controllable region~$\srcRegion_\T{C}$, \ie{}, the matrix~$\PORTmat$, defined in \eqref{eq:PortMatdef}, has non-zero entries only for indices~$p$ belonging to the edges that coincide with ports. Taking the smallest eigenvalue of~\eqref{eq:radEffoptim3}
\begin{equation}
\delta_\T{rad}^\T{lb} = \min\limits_i \left\{ \delta_i \right\}
\label{eq:radEffoptim4}
\end{equation}
and substituting it back into~\eqref{eq:radEffoptim1} yields the upper bound on radiation efficiency~$\eta_\T{rad}^\T{ub}$ for a given subregion~$\srcRegion_\T{C}$. The eigenvector of~\eqref{eq:radEffoptim3} corresponding to the smallest eigenvalue represents the optimal feeding.

As compared to the current density-based bounds, see, \eg{},~\cite{JelinekCapek_OptimalCurrentsOnArbitrarilyShapedSurfaces, GustafssonCapekSchab_TradeOffBetweenAntennaEfficiencyAndQfactor}, which are typically not reachable since full control of current is required, the bound~\eqref{eq:radEffoptim4} is realizable by impressing the corresponding voltage~$\Vport_i$ to selected ports. This implies that~\eqref{eq:radEffoptim4} is always sub-optimal with respect to the current-based bounds. Realistic antenna designs with a fixed placement of ports, such as those proposed in~\cite{2020_Kormilainen_etal_RealizingOptimalCurrentDistributions}, should rather be compared with~\eqref{eq:radEffoptim4} than with fundamental current density-based bounds~\cite{JelinekCapek_OptimalCurrentsOnArbitrarilyShapedSurfaces}.

\section*{Acknowledgement}
The authors would like to thank the two anonymous reviewers whose remarks improved the clarity and completeness of this paper.

\bibliographystyle{IEEEtran}
\bibliography{references}

\begin{IEEEbiography}[{\includegraphics[width=1in,height=1.25in,clip,keepaspectratio]{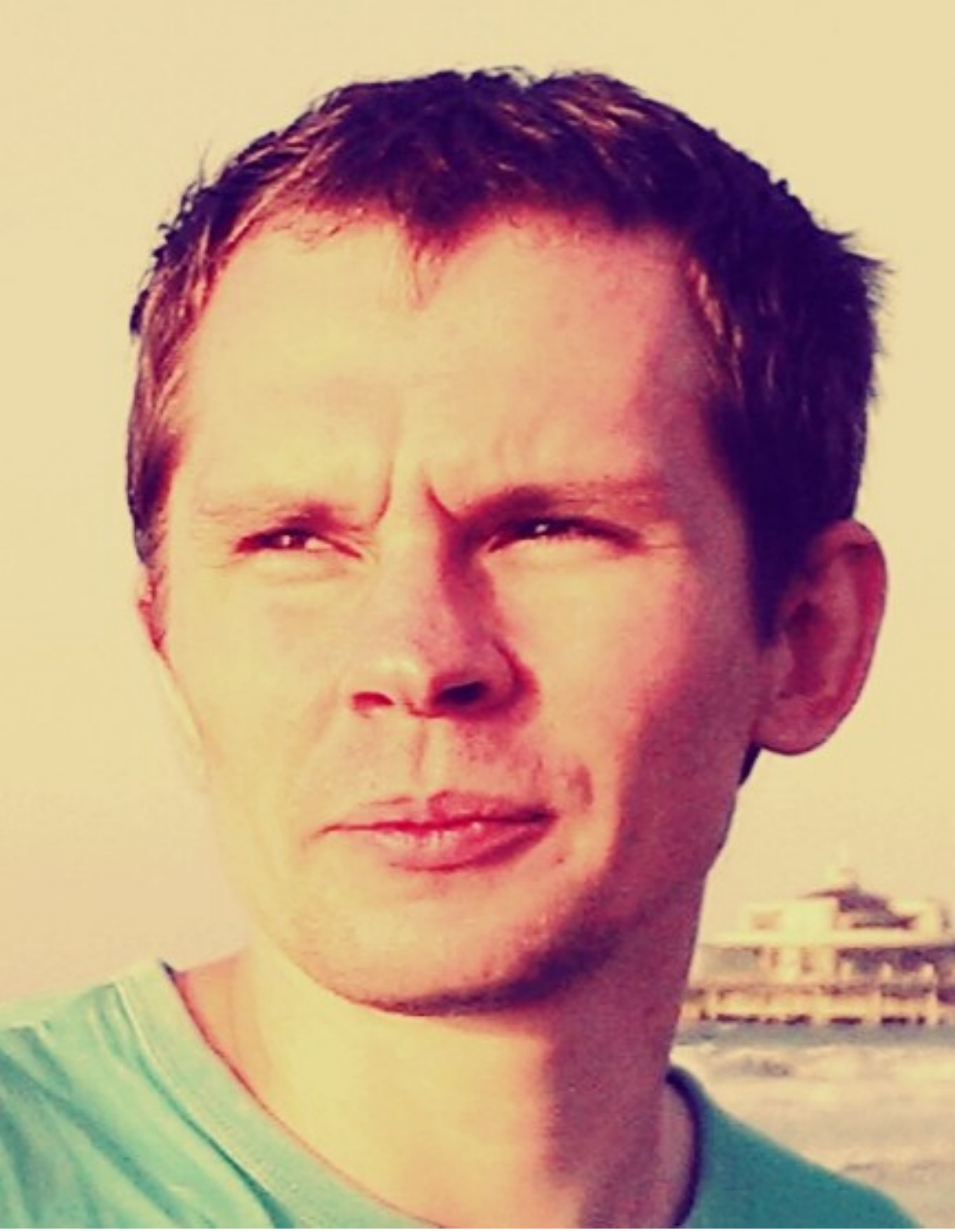}}]{Miloslav Capek}
(M'14, SM'17) received the M.Sc. degree in Electrical Engineering 2009, the Ph.D. degree in 2014, and was appointed Associate Professor in 2017, all from the Czech Technical University in Prague, Czech Republic.
	
He leads the development of the AToM (Antenna Toolbox for Matlab) package. His research interests are in the area of electromagnetic theory, electrically small antennas, numerical techniques, fractal geometry, and optimization. He authored or co-authored over 100~journal and conference papers.

Dr. Capek is member of Radioengineering Society, regional delegate of EurAAP, and Associate Editor of IET Microwaves, Antennas \& Propagation.
\end{IEEEbiography}

\begin{IEEEbiography}[{\includegraphics[width=1in,height=1.25in,clip,keepaspectratio]{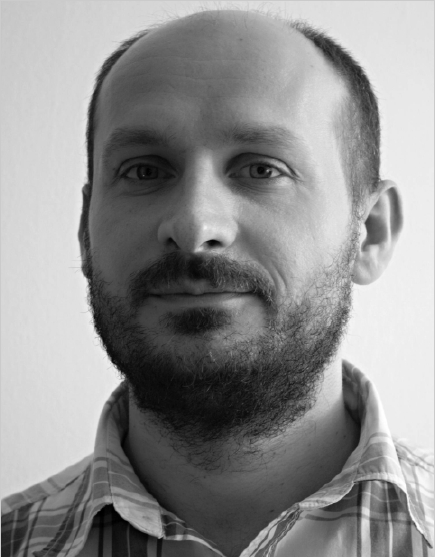}}]{Lukas Jelinek}
received his Ph.D. degree from the Czech Technical University in Prague, Czech Republic, in 2006. In 2015 he was appointed Associate Professor at the Department of Electromagnetic Field at the same university.

His research interests include wave propagation in complex media, general field theory, numerical techniques and optimization.
\end{IEEEbiography}

\begin{IEEEbiography}[{\includegraphics[width=1in,height=1.25in,clip,keepaspectratio]{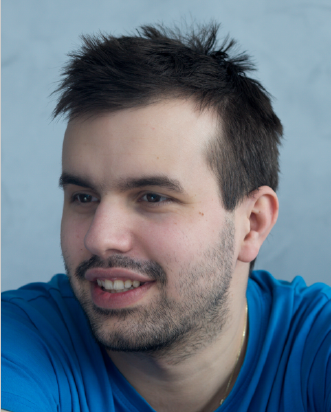}}]{Michal Masek}
received the M.Sc. degree in Electrical Engineering from Czech Technical University in Prague, Czech Republic, in 2015, where he is currently pursuing the Ph.D. degree in the area of modal tracking and characteristic modes. He is a member of the team developing the AToM (Antenna Toolbox for Matlab).
\end{IEEEbiography}

\end{document}